\documentclass[aps,prb,superscriptaddress,amsfonts,amsmath,amssymb,floatfix,showpacs,reprint,longbibliography,footinbib]{revtex4-2}%

\usepackage{graphicx} % Required for inserting images

\usepackage{svg}
\usepackage{amsmath,amssymb}
\usepackage{dcolumn}% Align table columns on decimal point
\usepackage{bm}% bold math
\usepackage{color}
\usepackage{hyperref}
\hypersetup{colorlinks=true,breaklinks,linkcolor=blue,urlcolor=blue,citecolor=blue}
\usepackage{xspace}
\usepackage{multirow}
\usepackage{physics}
\usepackage[version=4]{mhchem}
\usepackage{booktabs}
\usepackage[normalem]{ulem}

\newcommand{\prcdni}{\ce{PrCdNi4}}
\newcommand{\prmgni}{\ce{PrMgNi4}}
\newcommand{\prinni}{\ce{PrInNi4}}
\newcommand{\QW}{\bm{q}_\mathrm{W}}
\newcommand{\QWa}{\bm{q}_\mathrm{W1}}
\newcommand{\QWb}{\bm{q}_\mathrm{W2}}
\newcommand{\QWc}{\bm{q}_\mathrm{W3}}

\newcommand{\QXa}{\bm{q}_\mathrm{X1}}
\newcommand{\QXb}{\bm{q}_\mathrm{X2}}
\newcommand{\QXc}{\bm{q}_\mathrm{X3}}
\newcommand{\llangle}{\langle\!\langle}
\newcommand{\rrangle}{\rangle\!\rangle}
\newcommand{\dEjj}{\Delta E^{\textrm{($j$--$j$)}}}
\newcommand{\dELS}{\Delta E^{(LS)}}

\begin{document}

\title{Multipolar fluctuations in localized $4f^2$-electron systems \\ from dynamical mean-field theory: application to \prcdni}

\author{Koki Numa}
\affiliation{Department of Physics, Okayama University, Okayama 700-8530, Japan}
\author{Junya Otsuki}
\affiliation{Research Institute for Interdisciplinary Science, Okayama University, Okayama 700-8530, Japan}

\date{\today}

\begin{abstract}
Multipolar-fluctuation analysis based on density functional theory combined with dynamical mean-field theory is extended from $4f^1$- to $4f^2$-electron systems. Two methodological developments are presented. First, many-particle states are constructed within the $j$--$j$ coupling scheme in the $j=5/2$ subspace, so as to reproduce the crystalline-electric-field splitting of the $4f^2$ Hund’s-rule ground-state multiplet. Second, multipolar susceptibilities and interactions between $4f^2$ multiplets are derived through a mapping from single-particle to many-particle bases. We apply this framework to \prcdni, which has a non-Kramers $\Gamma_3$ doublet ground state, and propose a $(3z^2-r^2)$-type antiferroquadrupolar order at $\bm{q}=(2\pi,\pi,0)$ driven by the competition between nearest- and fourth-nearest-neighbor interactions. The formalism developed here is applicable to general $4f^n$-electron systems.
\end{abstract}

\maketitle

\section{Introduction}
\label{sec:introduction}

Rare-earth and actinide compounds with $4f$ or $5f$ electrons exhibit a variety of phase transitions at low temperatures arising from strong correlation effects~\cite{Kuramoto2009-review}. Their order parameters can be described systematically within a multipole framework that represents the anisotropy of the electronic internal degrees of freedom~\cite{Hayami2024-review}. Multipolar ordering gives rise to anomalous phenomena such as emergent chirality originating from electronic symmetry breaking in \ce{URhSn}~\cite{Ishitobi2026,Tokunaga2026}.

First-principles prediction of multipolar order remains a fundamental challenge in computational materials physics. The Ruderman–Kittel–Kasuya–Yosida (RKKY) interaction has been evaluated for real materials~\cite{Hanzawa2015-uru2si2-RKKY,Yamada2019-ceb6-RKKY,Iizuka2022-pr112-RKKY,Yamada2024-CeCoSi}.
Density functional theory combined with dynamical mean-field theory (DFT+DMFT) has been established as a powerful approach for strongly correlated materials~\cite{Georges1996,Kotliar2006}. Recent developments have enabled investigations of multipolar order within DFT+DMFT~\cite{Pourovskii2016,Otsuki2019-SCL}, and the resulting methods have been applied to specific compounds~\cite{Pourovskii2019,Pourovskii2020,Pourovskii2026}.
By analyzing multipolar fluctuations using the strong-coupling-limit formula~\cite{Otsuki2019-SCL}, we have successfully reproduced and predicted multipolar orders in CeB$_6$~\cite{Otsuki2024-ceb6}, CeRh$_6$Ge$_4$~\cite{itokazu2025-cerh6ge4}, and CeRh$_2$As$_2$~\cite{Numa2026-cerh2as2}.

In general, $4f^n$-electron systems with $n\ge 2$ host a rich variety of magnetic and multipolar phenomena.
Among them, Pr compounds with the $4f^2$ configuration exhibit distinctive behavior compared with Ce compounds. Particularly intriguing is the non-Kramers $\Gamma_3$ doublet realized in a cubic environment, which hosts quadrupolar degrees of freedom without magnetic dipoles~\cite{Hattori2014,Hattori2016,Ishitobi2021,Patri2019}. Quadrupolar ordering in the $\Gamma_3$ doublet system has been observed, e.g., in PrPb$_3$~\cite{Bucher1974-prpb3,Morin1982-prpb3,Tayama2001-prpb3,Onimaru2005-prpb3} and Pr$T_2X_{20}$ ($T$=Ir,~$X$=Zn;~$T$=V,~Ti,~$X$=Al)~\cite{Onimaru2010-prir2zn20,Sakai2011-prtr2al20,Tsujimoto2014-prv2al20,Sato2012-prti2al20,Onimaru2016-112review}. In addition, various physical properties have been reported, including ordinary Kondo behavior in \ce{PrMg3}~\cite{Tanida2006} and non-Fermi-liquid behavior seemingly associated with the quadrupolar Kondo effect~\cite{Onimaru2016-112review,Onimaru2016-nonfermi,Yamane2018-nonfermi}.

Applying the DFT+DMFT approach to $4f^n$ configurations with $n\geq 2$ requires special care.
At the heart of this issue lies the distinction between $LS$ coupling and $j$--$j$ coupling.
The $LS$ coupling first combines the orbital and spin angular momenta of multiple electrons separately as $\bm{L}=\sum_i\bm{l}_i$ and $\bm{S}=\sum_i\bm{s}_i$ under strong Coulomb interactions. The spin-orbit coupling then combines them to form the total angular momentum $\bm{J}=\bm{L}+\bm{S}$.
On the other hand, the $j$--$j$ coupling first combines the orbital and spin angular momenta of each electron as $\bm{j}_i=\bm{l}_i+\bm{s}_i$ under strong spin-orbit coupling. These single-particle angular momenta are then combined to form the total angular momentum $\bm{J}=\sum_i\bm{j}_i$.
The $LS$ coupling picture is useful for interpreting the crystalline-electric-field (CEF) states and magnetic properties governed by the $4f^n$ Hund's-rule ground multiplet.
By contrast, theoretical calculations such as DFT+DMFT are based on the single-particle Green's function and hence are compatible with the $j$--$j$ coupling.
Bridging the $j$--$j$ and $LS$ coupling pictures therefore constitutes a key step toward a consistent DFT+DMFT description of $4f^n$-electron systems.

In this paper, we extend the framework of multipolar-fluctuation analysis that has been applied to $4f^1$-electron systems to $4f^2$-electron systems. Specifically, we address the following two issues.
First, in Sec.~\ref{sec:CEF}, we discuss the relation between the one-body CEF potential and the CEF levels of the Hund's-rule ground multiplet.
This allows us to determine the CEF levels from the CEF potential estimated by DFT, and conversely, to adjust the CEF potential when necessary.
Second, in Sec.~\ref{sec:fn-fluctuation}, we address the momentum-dependent susceptibility, which is conventionally represented in the single-particle basis.
We develop a mapping from the single-particle susceptibility to the many-body susceptibility.
This enables direct analysis of multipolar fluctuations in the Hund's-rule ground multiplet.
Through application to the real material {\prcdni} in Sec.~\ref{sec:application}, we demonstrate its potential for first-principles exploration of multipolar physics in $4f^n$-electron systems.

\section{CEF levels within the $j$--$j$ coupling}
\label{sec:CEF}

CEF levels play important roles in the magnetic and, more generally, multipolar properties of localized $4f$ electrons.
In this section, 
we discuss the CEF levels of the Hund's-rule ground-state multiplet ($J=4$) of the $4f^2$ configuration, calculated as the eigenstates of a many-body Hamiltonian constructed within the $j$--$j$ coupling scheme.
The focus is placed on how the so-called Lea--Leask--Wolf diagram~\cite{Lea1962-LLW} can be reproduced within the $j=5/2$ subspace, neglecting the high-energy $j=7/2$ states. 

\begin{figure*}[!t]
      \includegraphics[width=0.4\linewidth]{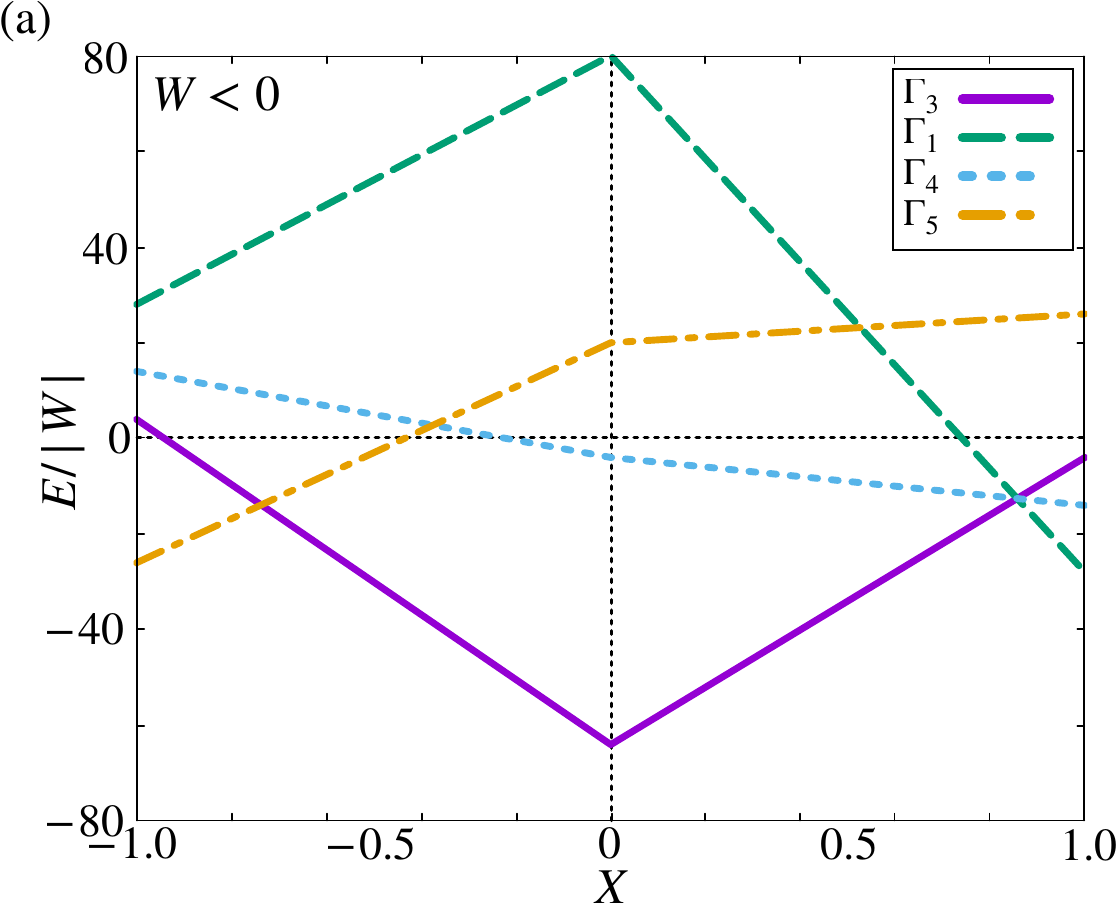}
      \includegraphics[width=0.4\linewidth]{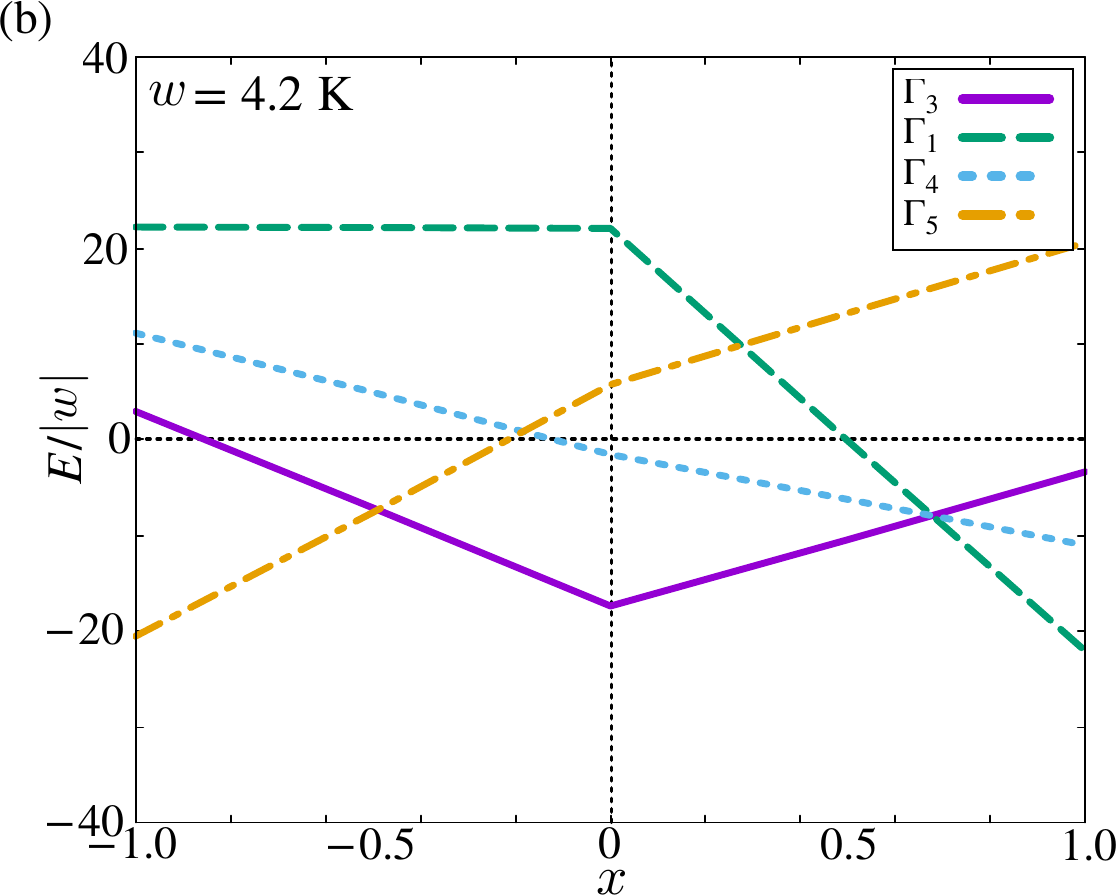}
      \includegraphics[width=0.4\linewidth]{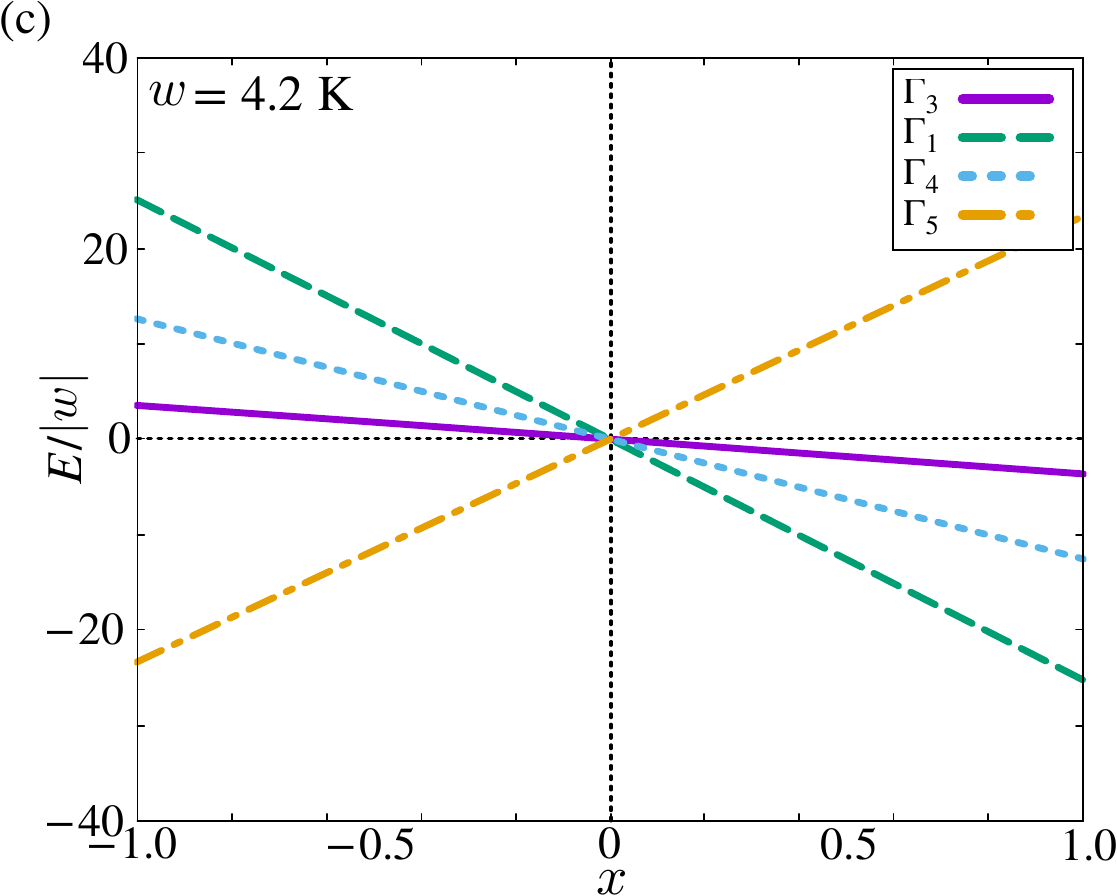}
      \includegraphics[width=0.4\linewidth]{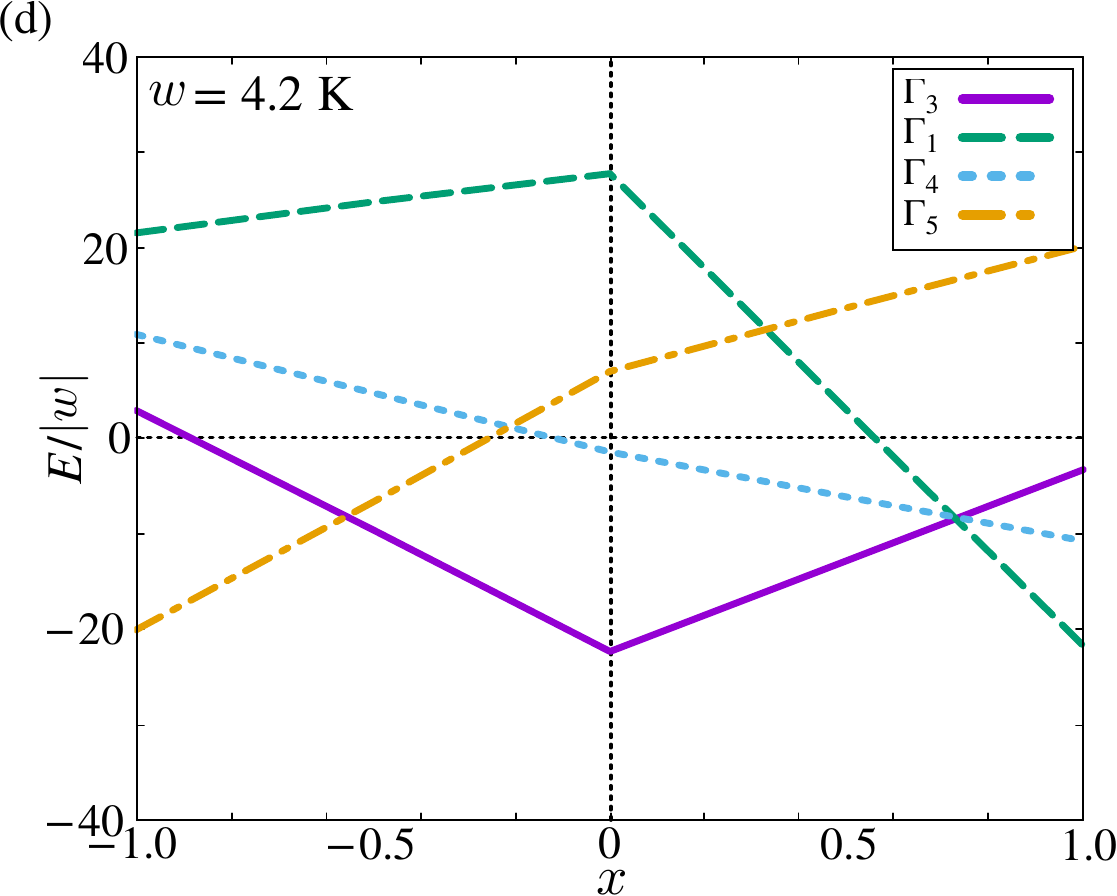}
      \caption{CEF levels within the $J=4$ manifold of the $4f^2$ configuration obtained from (a) the Hamiltonian in the $LS$ coupling $\hat{H}^{(f^2)}_{\mathrm{CEF}}$ given by Eq.~\eqref{eq:f2-CEF}, (b) the many-body Hamiltonian $\mathcal{H}$ in the full $l=3$ space including both the $j=5/2$ and $j=7/2$ orbitals formulated in Eqs.~\eqref{eq:many-Hamiltonian}--\eqref{eq:f1many-int}, (c) the effective $j=5/2$ Hamiltonian $\mathcal{H}^{(5/2)}_{\mathrm{eff}}$ introduced in Eqs.~\eqref{eq:effjj-H}--\eqref{eq:effjj-int-matele}, retaining only the zeroth-order terms corresponding to the conventional $j$--$j$ coupling, and (d) $\mathcal{H}^{(5/2)}_{\mathrm{eff}}$ incorporating the first-order corrections in $1/\lambda$. For (b)--(d), the Coulomb interaction parameters are taken as $F^0=5.7~\mathrm{eV},~F^2=10~\mathrm{eV},~F^4=6.7~\mathrm{eV},$ and $F^6=4.9~\mathrm{eV}$, with the spin-orbit coupling constant $\lambda=0.2~\mathrm{eV}$ and the CEF scaling parameter $w=4.2~\mathrm{K}$. A detailed description of the choice of these parameters is given in Sec.~\ref{subsec:prcdni4-DMFT}.
      }
      \label{fig:LLW}
\end{figure*}

CEF levels of $4f^2$-electron systems in a cubic environment are obtained by diagonalizing the Hamiltonian within the $J=4$ Hund's-rule ground-state manifold given by~\cite{Lea1962-LLW}
\begin{align}
      \label{eq:f2-CEF}
      \hat{H}_{\mathrm{CEF}}^{(f^2)}=W\Bigg[& X\frac{\hat{O}_4^0(4)+5\hat{O}_4^4(4)}{60}
      \notag \\
      &+\left(1-|X|\right)\frac{\hat{O}_6^0(4)-21\hat{O}_6^4(4)}{1260}\Bigg].
\end{align}
Here, $\hat{O}_p^q(J)$ are the Stevens operator equivalents \cite{Stevens1952,Hutching1964}, where the hat indicates a matrix in the $J$ space. $W$ sets the overall CEF energy scale and $X$ parametrizes the relative weight of the $\hat{O}_4^q$ and $\hat{O}_6^q$ terms. Figure~\ref{fig:LLW}(a) shows the CEF levels for the $J=4$ manifold with respect to $X$. The non-Kramers $\Gamma_3$ doublet ground state is realized in the range $-0.73 < X < 0.85$ for $W<0$.

On the other hand, theoretical analyses typically employ single-particle states and introduce Coulomb interactions afterward. The CEF Hamiltonian for the angular momentum $l=3$ is given by~\cite{Lea1962-LLW}
\begin{align}
      \label{eq:f1-CEF}
      \hat{H}_{\mathrm{CEF}}^{(f^1)}=w\Bigg[& x\frac{\hat{O}_4^0(3)+5\hat{O}_4^4(3)}{15}
      \notag \\
      &+\left(1-|x|\right)\frac{\hat{O}_6^0(3)-21\hat{O}_6^4(3)}{180}\Bigg].
\end{align}
Note that the CEF parameters $w$ and $x$ in Eq.~\eqref{eq:f1-CEF} are distinct from the aforementioned $W$ and $X$ in Eq.~\eqref{eq:f2-CEF}. The CEF levels of the $J=4$ multiplet in the $4f^2$ configuration are obtained as the eigenvalues of the many-body Hamiltonian given by~\cite{Hotta2003-localmodel,Hotta2006}
\begin{align}
      \label{eq:many-Hamiltonian}
      \mathcal{H}  & =\mathcal{H}_{\mathrm{SOC}}+\mathcal{H}^{(f^1)}_{\mathrm{CEF}}+\mathcal{H}_{\mathrm{int}},
\end{align}
with
\begin{align}
      \label{eq:f1many-SOC}
      \mathcal{H}_{\mathrm{SOC}} & =\lambda\sum_{mm'}\sum_{\sigma\sigma'}f^{\dag}_{m\sigma}(\bm{l}_{mm'}\cdot\bm{s}_{\sigma\sigma'})f_{m'\sigma'}, \\
      \label{eq:f1many-CEF}
      \mathcal{H}^{(f^1)}_{\mathrm{CEF}} & =\sum_{mm'}\sum_{\sigma}B_{mm'}f^{\dag}_{m\sigma}f_{m'\sigma}, \\
      \label{eq:f1many-int}
      \mathcal{H}_{\mathrm{int}} & =\sum_{m_1m_2m_3m_4}\sum_{\sigma\sigma'}I_{m_1m_2,m_3m_4}f^{\dag}_{m_1\sigma}f^{\dag}_{m_2\sigma'}f_{m_3\sigma'}f_{m_4\sigma},
\end{align}
where $f_{m\sigma}$ is the annihilation operator for an $f$ electron, and $m$ and $\sigma$ are the $z$ components of the orbital angular momentum ($l=3$) and the spin ($s=1/2$), respectively. The first term in Eq.~\eqref{eq:many-Hamiltonian} is the spin-orbit coupling with the coupling constant $\lambda$, which yields the energy levels $\lambda_a=-2\lambda$ for the $j=5/2$ states and $\lambda_b=(3/2)\lambda$ for the $j=7/2$ states. The second term represents the CEF potential, where the matrix elements are given by $B_{mm'} = (\hat{H}^{(f^1)}_{\mathrm{CEF}})_{mm'}$. The final term denotes the $f$--$f$ Coulomb interaction, whose matrix elements $I_{m_1m_2,m_3m_4}$ are parametrized by the Slater integrals $F^0, F^2, F^4$, and $F^6$ for $l=3$~\cite{Anisimov1997}.
 
Figure~\ref{fig:LLW}(b) shows the corresponding CEF levels obtained by diagonalizing Eqs.~\eqref{eq:many-Hamiltonian}--\eqref{eq:f1many-int} in the full $l=3$ space ($j=5/2$ and $j=7/2$). The overall behavior of the CEF splitting reproduces the result of the $LS$ coupling in Fig.~\ref{fig:LLW}(a).

In complicated many-body calculations, it is conventional to restrict the analysis to the $j=5/2$ subspace.
Following Ref.~\cite{Hotta2006}, we incorporate the effect of the $j=7/2$ states by perturbation theory with respect to $1/\lambda$.
To this end, we define the terms other than the spin-orbit coupling in Eq.~\eqref{eq:many-Hamiltonian} as
\begin{align}
      \label{eq:jj-pert}
      \mathcal{H}' = \mathcal{H}^{(f^1)}_{\mathrm{CEF}} + \mathcal{H}_{\mathrm{int}}.
\end{align}
Denoting the projection operator onto the $j=5/2$ subspace by $\mathcal{P}$, the zeroth-order term is given by $\mathcal{P}\mathcal{H}’\mathcal{P}$.
Thus, the remaining terms in $\mathcal{H}’$ give rise to perturbative corrections.
The effective Hamiltonian $\mathcal{H}^{(5/2)}_{\mathrm{eff}}$ defined within the $j=5/2$ subspace is, up to first order in $1/\lambda$, given by~\cite{Hotta2006}
\begin{align}
      \label{eq:effjj-H}
      \mathcal{H}^{(5/2)}_{\mathrm{eff}} & =\mathcal{H}^{(5/2)}_{\mathrm{CEF}}+\mathcal{H}^{(5/2)}_{\mathrm{int}},                                                                                                                                                   \\
      \label{eq:effjj-CEF}
      \mathcal{H}^{(5/2)}_{\mathrm{CEF}} & =\sum_{\alpha_1\alpha_2}\left(\tilde{B}^{(0)}_{\alpha_1\alpha_2}+\tilde{B}^{(1)}_{\alpha_1\alpha_2}\right)f^{\dag}_{\alpha_1}f_{\alpha_2},                                                                                \\
      \label{eq:effjj-int}
      \mathcal{H}^{(5/2)}_{\mathrm{int}} & =\sum_{\alpha_1\alpha_2\alpha_3\alpha_4}\left(\tilde{I}^{(0)}_{\alpha_1\alpha_2,\alpha_3\alpha_4}+\tilde{I}^{(1)}_{\alpha_1\alpha_2,\alpha_3\alpha_4}\right)f^{\dag}_{\alpha_1}f^{\dag}_{\alpha_2}f_{\alpha_3}f_{\alpha_4},
\end{align}
where $\alpha$ represents the $z$ component of the total angular momentum ($j=5/2$). The first and second terms in Eq.~\eqref{eq:effjj-H} denote the CEF potential and the Coulomb interaction within the $j=5/2$ subspace, respectively. In Eqs.~\eqref{eq:effjj-CEF} and \eqref{eq:effjj-int}, $\tilde{B}^{(0)}$ and $\tilde{I}^{(0)}$ represent the zeroth-order terms in the perturbation expansion, while $\tilde{B}^{(1)}$ and $\tilde{I}^{(1)}$ correspond to the first-order corrections.

The zeroth-order matrix elements $\tilde{B}^{(0)}$ and $\tilde{I}^{(0)}$ in Eqs.~\eqref{eq:effjj-CEF} and \eqref{eq:effjj-int} are obtained by transforming the CEF potential and Coulomb interaction terms of Eq.~\eqref{eq:jj-pert} into the total-angular-momentum basis and restricting them to the $j=5/2$ subspace. These terms correspond to the conventional $j$--$j$ coupling in the limit of infinitely large spin-orbit coupling ($\lambda\to\infty$).

Figure~\ref{fig:LLW}(c) shows the CEF levels obtained from Eqs.~\eqref{eq:effjj-H}--\eqref{eq:effjj-int} with only the zeroth-order CEF and interaction terms retained. Because the contributions from the $j=7/2$ orbitals are neglected, this result fails to reproduce the $LS$-coupling CEF levels in Fig.~\ref{fig:LLW}(a). In particular, the non-Kramers $\Gamma_3$ doublet can never be the ground state.

The first-order correction in $1/\lambda$ to the CEF potential, $\tilde{B}^{(1)}$ in Eq.~\eqref{eq:effjj-CEF}, is given by~\cite{Hotta2006}
\begin{align}
      \label{eq:effjj-CEF-matele}
      \tilde{B}^{(1)}_{\alpha_1\alpha_2}= & \sum_{\beta}\frac{\bra{\alpha_1}\mathcal{H}^{(f^1)}_{\mathrm{CEF}}\ket{\beta}\bra{\beta}\mathcal{H}^{(f^1)}_{\mathrm{CEF}}\ket{\alpha_2}}{\lambda_a-\lambda_b},
\end{align}
where $\beta$ labels the $z$ component of the total angular momentum for the $j=7/2$ states, $\ket{\alpha}=f^{\dag}_{\alpha}\ket{0}$, and $\ket{\beta}=f^{\dag}_{\beta}\ket{0}$. The matrix elements of the Coulomb interaction $\tilde{I}^{(1)}$ in Eq.~\eqref{eq:effjj-int} are given by
\begin{align}
      \label{eq:effjj-int-matele}
      \tilde{I}^{(1)}_{\alpha_1\alpha_2,\alpha_3\alpha_4}= & \sum_{\alpha\beta}\frac{\bra{\alpha_1\alpha_2}\mathcal{H}'\ket{\alpha\beta}\bra{\beta\alpha}\mathcal{H}'\ket{\alpha_3\alpha_4}}{\lambda_a-\lambda_b} \notag              \\
                                                           & +\sum_{\beta_1\beta_2}\frac{\bra{\alpha_1\alpha_2}\mathcal{H}'\ket{\beta_1\beta_2}\bra{\beta_2\beta_1}\mathcal{H}'\ket{\alpha_3\alpha_4}}{2(\lambda_a-\lambda_b)} \notag \\
                                                           & -\bra{\alpha_1\alpha_2}\sum_{\alpha\alpha'}\tilde{B}^{(1)}_{\alpha\alpha'}f^{\dag}_{\alpha}f_{\alpha'}\ket{\alpha_3\alpha_4},
\end{align}
where
$\ket{\alpha_1\alpha_2}=f^{\dag}_{\alpha_1}f^{\dag}_{\alpha_2}\ket{0}$, $\ket{\alpha\beta}=f^{\dag}_{\alpha}f^{\dag}_{\beta}\ket{0}$, and $\ket{\beta_1\beta_2}=f^{\dag}_{\beta_1}f^{\dag}_{\beta_2}\ket{0}$. The final term in Eq.~\eqref{eq:effjj-int-matele} is included to avoid double counting of the first-order CEF contributions in $1/\lambda$.

Figure~\ref{fig:LLW}(d) shows the corresponding CEF levels from Eqs.~\eqref{eq:effjj-H}--\eqref{eq:effjj-int} with both the zeroth- and first-order terms retained.
This effective $j=5/2$ model reproduces the $LS$-coupling results in Fig.~\ref{fig:LLW}(a) owing to the inclusion of the contributions from the $j=7/2$ orbitals as a perturbation. This correction allows the non-Kramers $\Gamma_3$ doublet to become the ground state.

Finally, we briefly comment on the validity of the perturbation expansion. The first-order correction $\tilde{I}^{(1)}$ to the CEF splitting induced by the $j=7/2$ orbitals scales as $|w|U/\lambda$. Since this correction must be sufficiently smaller than the scale $U$ of the corresponding zeroth-order term $\tilde{I}^{(0)}$, the validity of the perturbation expansion requires $|w|/\lambda\ll 1$. This condition is fulfilled in typical $f$-electron materials because $|w| \sim 0.001$--$0.01$~eV and $\lambda \sim 0.2$~eV. The parameters estimated in Sec.~\ref{subsec:prcdni4-DMFT} indeed satisfy this requirement.

\section{Many-body susceptibility}
\label{sec:fn-fluctuation}

The conventional static susceptibility in the single-particle basis is given by
\begin{align}
      \label{eq:f1chi-def}
      \chi_{m_1m_2,m_3m_4} & = \int_0^{\beta} d\tau\langle f^{\dag}_{m_1}(\tau)f_{m_2}(\tau)f^{\dag}_{m_4}f_{m_3}\rangle,
\end{align}
where $m$ denotes a general single-particle state, e.g., $m=\alpha$ for the $j=5/2$ states in Sec.~\ref{sec:CEF}.
With $(m_1,m_2)$ treated as a composite index, $\chi$ in Eq.~\eqref{eq:f1chi-def} can be regarded as an $N_m^2\times N_m^2$ matrix, where $N_m$ is the number of the single-particle states. On the other hand, the susceptibility expressed in the many-particle basis $M$ is given by
\begin{align}
      \label{eq:f2chi-def}
      \mathcal{X}_{M_1M_2,M_3M_4}=\int_0^{\beta} d\tau\langle X_{M_1M_2}(\tau)X_{M_3M_4}\rangle,
\end{align}
where $X_{MM'}=\ket{M}\bra{M'}$ represents the transition operator from $M'$ to $M$. Similarly, with $(M_1,M_2)$ treated as a composite index, $\mathcal{X}$ in Eq.~\eqref{eq:f2chi-def} forms an $N_M^2\times N_M^2$ matrix, where $N_M$ is the number of the many-particle states.

Fluctuations of multipoles within the Hund's-rule ground-state multiplet are described by $\mathcal{X}$ in Eq.~\eqref{eq:f2chi-def}.
Its momentum dependence can be used to detect phase transitions into multipolar ordered states.
On the other hand, a methodology within DFT+DMFT has been established for $\chi$ in Eq.~\eqref{eq:f1chi-def}. 
The momentum-dependent $\chi$ can be computed through the Bethe-Salpeter equation~\cite{Georges1996}, and its practical approximate formula has also been developed~\cite{Otsuki2019-SCL}.
We therefore pursue a conversion from $\chi$ to $\mathcal{X}$ in this section.
We note that intersite interactions in the $M$ basis have also been formulated in combination with a local approximation~\cite{Pourovskii2016,Savrasov2004}.
The conversion method presented below provides a general connection between quantities represented in single-particle and many-particle bases.

We relate the required susceptibility $\mathcal{X}_{M_1M_2,M_3M_4}$ to the conventional one $\chi_{m_1m_2,m_3m_4}$. Inserting the complete set of many-particle states $M$ into Eq.~\eqref{eq:f1chi-def}, we obtain
\begin{align}
      \label{eq:f2-to-f1chi}
      \chi_{m_1m_2,m_3m_4}
        =& \sum_{M_1M_2M_3M_4} A_{m_1m_2,M_1M_2} \notag                        \\
       & \times \mathcal{X}_{M_1M_2,M_3M_4} A^{\dag}_{M_3M_4,m_3m_4},
\end{align}
where $A$ is the matrix element of the density operator with respect to the many-particle state $M$, defined by
\begin{align}
      \label{eq:A-matrix}
      A_{mm', MM'} = \bra{M} f^{\dag}_{m} f_{m'} \ket{M'}.
\end{align}
$A$ can be regarded as a conversion matrix of size $N_m^2\times N_M^2$.
How to evaluate $A$ is presented in Appendix~\ref{app:A-matrix}.

The summation in Eq.~\eqref{eq:f2-to-f1chi} runs over all many-particle states with different particle numbers. However, contributions from high-energy states are negligible because of the Boltzmann factor in the thermal average. This allows us to restrict the summation to low-energy states, such as the $J=4$ Hund's-rule ground multiplet for the $4f^2$-electron system. Consequently, Eq.~\eqref{eq:f2-to-f1chi} is approximated by
\begin{align}
      \label{eq:strict-J4}
        \chi_{m_1m_2,m_3m_4} \simeq& \sideset{}{'}\sum_{M_1M_2M_3M_4}A_{m_1m_2,M_1M_2} \notag    \\
       & \times \mathcal{X}_{M_1M_2,M_3M_4} A^{\dag}_{M_3M_4,m_3m_4},
\end{align}
where the prime on the summation denotes the restriction to the low-energy states.

Equation~\eqref{eq:strict-J4} provides the forward transformation that expresses $\chi_{m_1m_2,m_3m_4}$ in terms of $\mathcal{X}_{M_1M_2,M_3M_4}$. Conversely, we perform an inverse transformation from $\chi_{m_1m_2,m_3m_4}$ to $\mathcal{X}_{M_1M_2,M_3M_4}$ as
\begin{align}
      \label{eq:f1-to-f2chi}
      \mathcal{X}_{M_1M_2,M_3M_4}\simeq & \sum_{m_1m_2m_3m_4}A^+_{M_1M_2,m_1m_2} \notag                        \\
                                 & \times\chi_{m_1m_2,m_3m_4}\left(A^{+}\right)^{\dag}_{m_3m_4,M_3M_4},
\end{align}
where $A^+$ is the Moore--Penrose pseudoinverse of the matrix $A$.
The idea of this transformation is as follows.
For simplicity, we express $\chi$ and $\mathcal{X}$ in Eqs.~\eqref{eq:f1chi-def} and \eqref{eq:f2chi-def} symbolically as $\chi = \llangle \bm{x}; \bm{x}^{\dag} \rrangle$ and $\mathcal{X} = \llangle \bm{X}; \bm{X}^{\dag} \rrangle$, respectively.
Here, $\bm{x}$ and $\bm{X}$ denote vectors representing $f_{m}^{\dag}f_{m'}$ and $X_{MM'}$, respectively.
Equation~\eqref{eq:strict-J4} indicates a formal relation of the form $\bm{x}=A\bm{X}$.
The corresponding inverse transformation for $\bm{X}$ is constructed with the Moore--Penrose pseudoinverse as $\tilde{\bm{X}}=A^{+}\bm{x}$.
Thus, we obtain Eq.~\eqref{eq:f1-to-f2chi}.

The meaning of $\tilde{\bm{X}}=A^{+}\bm{x}$ depends on the shape of $A$. Let the size of $A$ be $\mu\times\nu$: The linear system $\bm{x}=A\bm{X}$ consists of $\mu$ equations for $\nu$ variables. The case $\mu>\nu$ corresponds to an overdetermined problem, for which the pseudoinverse is given by
\begin{align}
      \label{eq:over-A}
      A^+=\left(A^{\dag}A\right)^{-1}A^{\dag}.
\end{align}
In this case, $\tilde{\bm{X}}=A^{+}\bm{x}$ yields the least-squares solution that minimizes $\|A\bm{X}-\bm{x}\|^2$, where $\| \cdot \|$ stands for the standard vector norm ($L_2$ norm). For an underdetermined problem with $\mu<\nu$, on the other hand, the pseudoinverse is given by
\begin{align}
      \label{eq:under-A}
      A^+=A^{\dag}\left(AA^{\dag}\right)^{-1}.
\end{align}
In this case, $\tilde{\bm{X}}=A^{+}\bm{x}$ yields the solution of the $L_2$-norm minimization, which minimizes $\|\bm{X}\|^2$ subject to $A\bm{X}=\bm{x}$~\cite{Otsuki2020-sparse,Gentle2007}. 

In practice, the number of independent equations in $\bm{x}=A\bm{X}$ is limited by the rank of $A$, $\text{rank}(A)$. When $\text{rank}(A) < \min(\mu, \nu)$, i.e., when $A$ is rank deficient,  dimensional reduction using singular value decomposition is required for evaluating Eqs.~\eqref{eq:over-A} and \eqref{eq:under-A}~\cite{Press2007}.

Whether the problem is overdetermined or underdetermined depends on the rare-earth ion and low-energy states considered.
For example, in the $4f^2$ configuration of Pr$^{3+}$, we can restrict the low-energy states to the $J=4$ multiplet, neglecting the other multiplets at energies higher than about 1~eV. The dimension $\nu$ (the number of $MM'$ pairs) is then $\nu=9^2=81$.
For the single-particle states, we consider either only the $j=5/2$ states or both the $j=5/2$ and $j=7/2$ states. The corresponding dimension $\mu$ (the number of $mm’$ pairs) is $\mu=6^2=36$ or $\mu=14^2=196$, respectively.
The former case, with $\mu<\nu$, is underdetermined.
If the $j=7/2$ states are included, $\mu=196$ apparently suggests an overdetermined problem. However, because of the rank deficiency, the number of independent equations is reduced to $\mathrm{rank}(A)=64$. Since $\mathrm{rank}(A)<\nu$, the problem remains underdetermined.
There is, however, a possibility of recovering an overdetermined condition by restricting $M$ to low-lying CEF states, which is justified at low temperatures.
See Appendix~\ref{app:over-under} for general $4f^n$ configurations.

\section{Application to a real material: Quadrupolar order in \prcdni}
\label{sec:application}

In the preceding sections, we extended the theoretical framework to $4f^2$-electron systems by constructing the effective CEF potential within the $j=5/2$ subspace and formulating the susceptibility in the many-particle basis. In this section, we apply this approach to \prcdni~as a Pr compound featuring a non-Kramers $\Gamma_3$ doublet ground state.

\subsection{Physical properties from experiments}
\label{subsec:properties}

\begin{figure}[t]
      \includegraphics[width=1.0\linewidth]{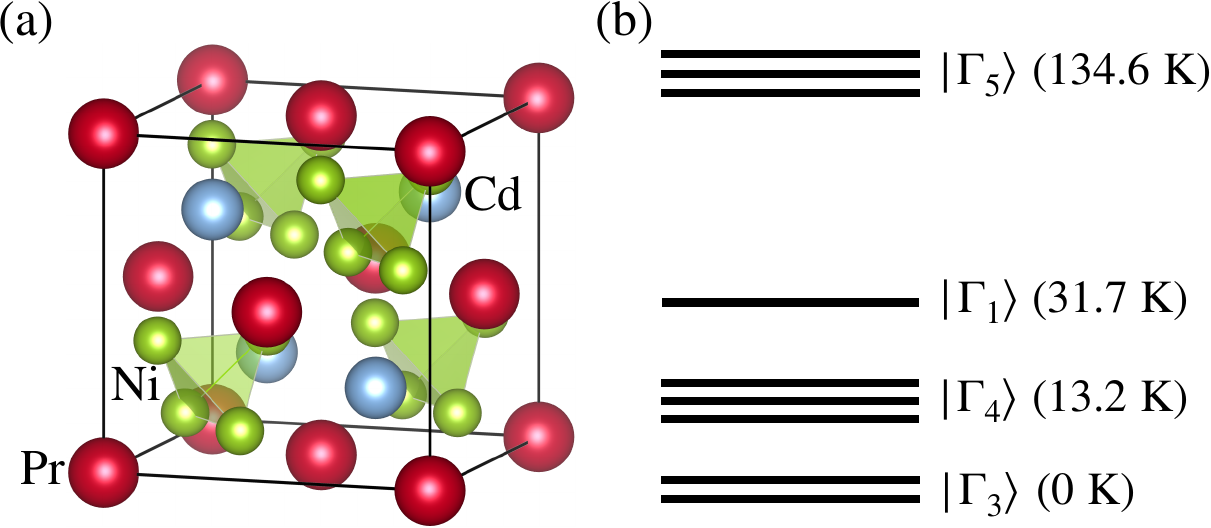}
      \caption{(a) Crystal structure of \prcdni. (b) The CEF level scheme for the $J=4$ manifold of the $4f^2$ configuration derived from experiments~\cite{Kusanose2025-prcdni4,Kusanose2022-neutron}, corresponding to the parameters $W=-3\;\mathrm{K}$ and $X=0.8$ in Eq.~\eqref{eq:f2-CEF}. The crystal structure was drawn using VESTA \cite{VESTA}.
      }
      \label{fig:str}
\end{figure}

\prcdni~has the $\mathrm{MgSnCu}_4$-type crystal structure, belonging to the space group $F\bar{4}3m$ ($T_d^2$, No. 216), where the Pr atoms are located at sites with $T_d$ site symmetry~\cite{Tappe2012-recdni4,Kusanose2025-prcdni4}. This structure is classified as a ternary ordered Laves phase, in which the Pr atoms form a face-centered-cubic (fcc) lattice as illustrated in Fig.~\ref{fig:str}(a). This family of Pr-based compounds exhibits a wide variety of physical properties, such as the quadrupolar order hindered by atomic disorder in \prmgni~\cite{Kusanose2019-prmgni4,Kusanose2019-prmgni4-mag,Kusanose2022-neutron} and the field-induced ferromagnetic transition in \prinni~\cite{Tsujii2002-prinni4,Suzuki2003-prinni4-gs, walker2006-prinni4-ins}. Consequently, the interplay between their structural features and various physical properties has attracted intense interest.

\prcdni~undergoes a nonmagnetic phase transition at $T_{\mathrm{O}}=1.0~\mathrm{K}$, detected in specific-heat and resistivity measurements~\cite{Kusanose2025-prcdni4}. The $4f^2(J=4)$ CEF levels have been identified as illustrated in Fig.~\ref{fig:str}(b), with the non-Kramers $\Gamma_3$ doublet as the ground state and the first-excited $\Gamma_4$ triplet separated by $13.2~\mathrm{K}$. This level scheme corresponds to the CEF parameters $W=-3\;\mathrm{K}$ and $X=0.8$ in Eq.~\eqref{eq:f2-CEF}. Since this non-Kramers doublet possesses no magnetic-dipole degrees of freedom, the nonmagnetic transition at $T=T_{\mathrm{O}}$ is expected to be of antiferroquadrupolar origin associated with the $\Gamma_3$ degrees of freedom.

\subsection{DFT calculations}
\label{subsec:prcdni4-DFT}

\begin{figure}[t]
      \includegraphics[width=1.0\linewidth]{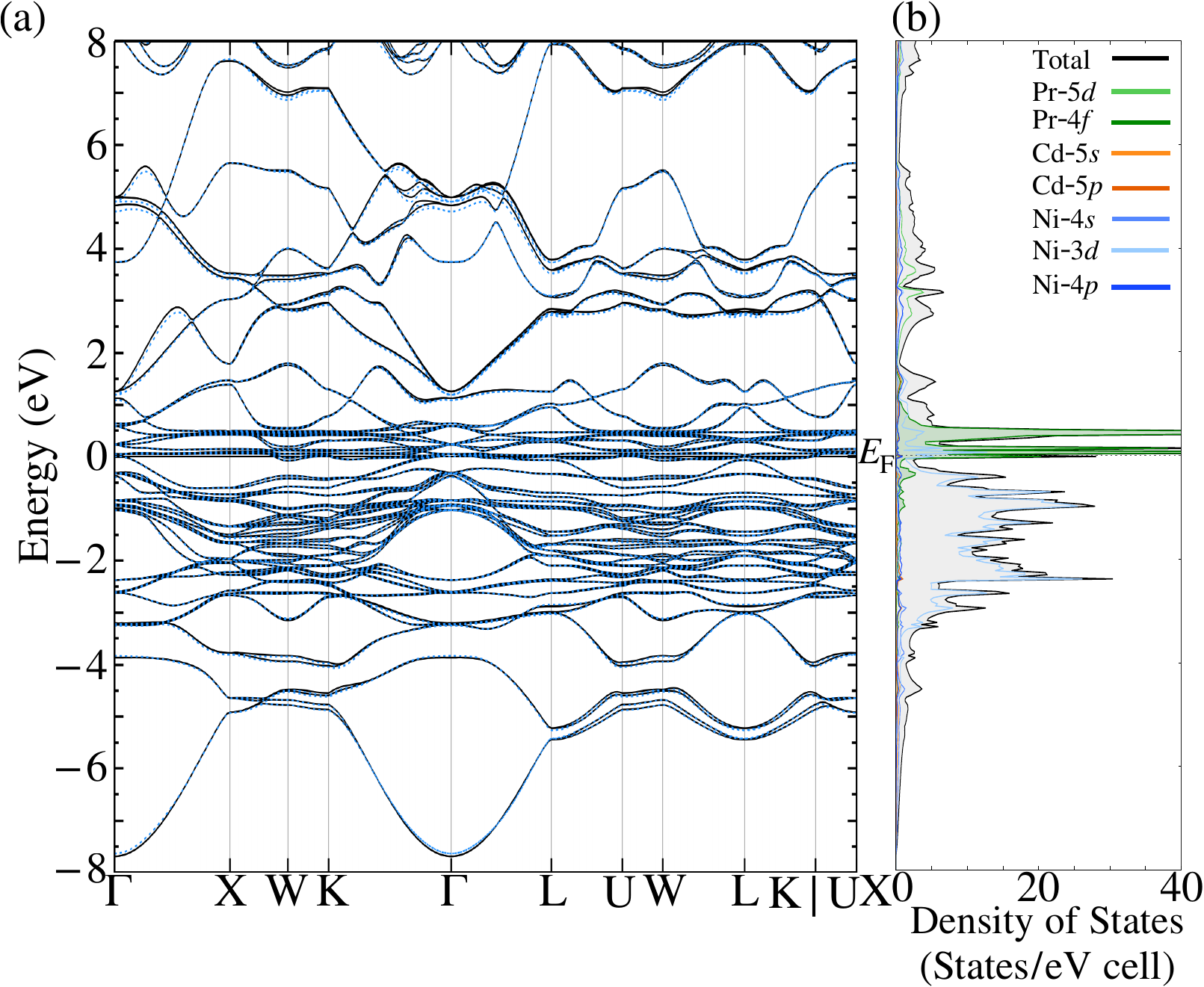}
      \includegraphics[width=0.45\linewidth]{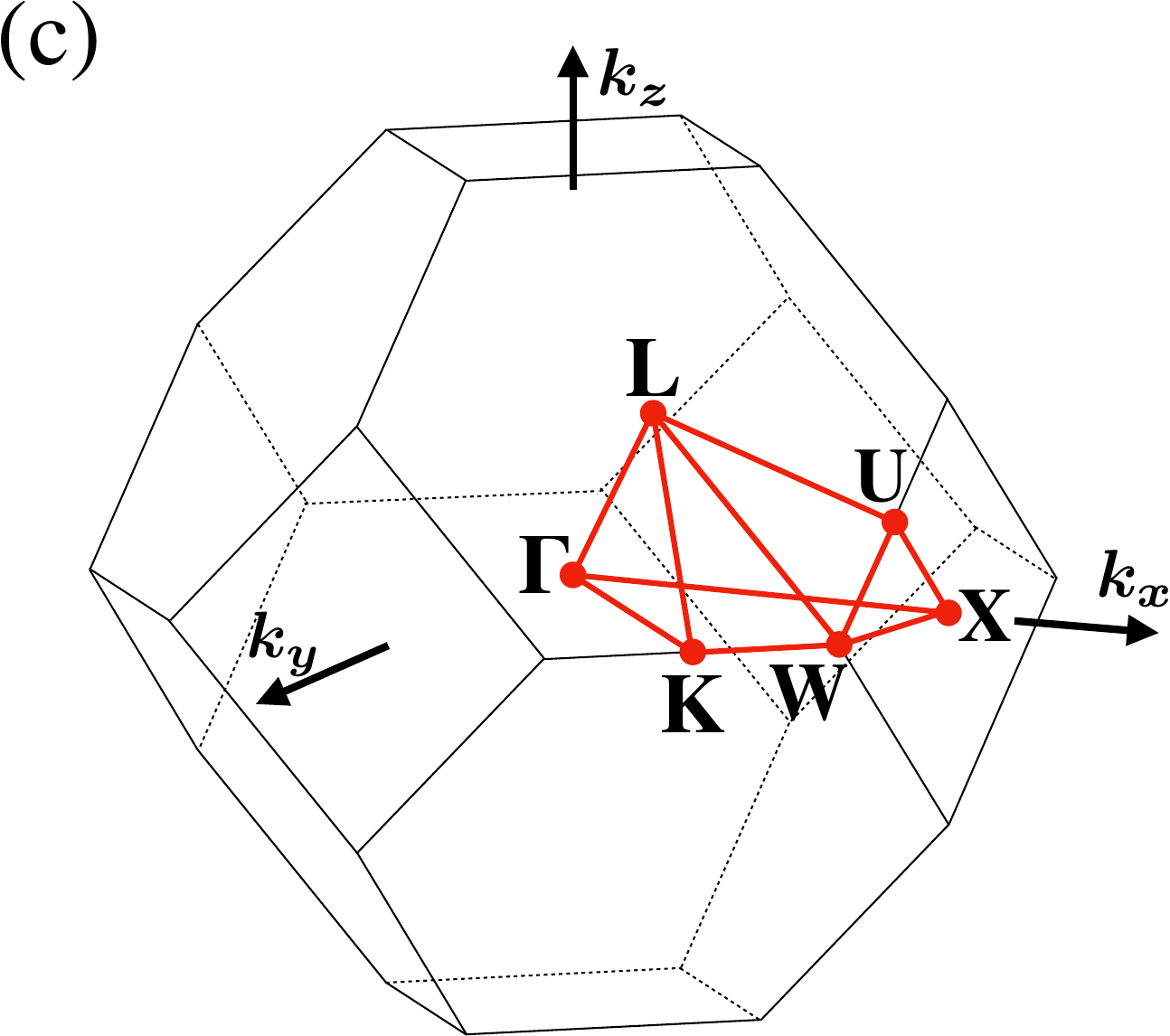}
      \caption{(a) Fully relativistic band structure of \prcdni~calculated using the FPLO package. The black lines represent the DFT result and the blue dashed lines represent the tight-binding bands obtained from projective Wannier functions. (b) Density of states. The black and colored lines show the total and orbital-resolved contributions, respectively. (c) The high-symmetry points in the Brillouin zone. The red line indicates the path used in (a) and subsequent plots.}
      \label{fig:band_DFT}
\end{figure}

In this calculation scheme, we first compute the electronic structure of \prcdni~based on DFT, treating the Pr $4f$ electrons as itinerant. The structural parameters are provided in Ref.~\cite{Kusanose2025-prcdni4}. The calculations are performed using the full-potential local-orbital (FPLO) package~\cite{Koepernik1999-FPLO,Opahle1999-FPLO} with a fully relativistic treatment, employing the generalized gradient approximation (PBE96) for the exchange-correlation potential~\cite{Perdew1996-GGA}. A $24 \times 24 \times 24$ $k$-point mesh is used for sampling in the full Brillouin zone. Figure~\ref{fig:band_DFT} shows the DFT band structure and density of states. Since this material crystallizes in the noncentrosymmetric space group $F\bar{4}3m$, the bands split as a consequence of parity violation. The contribution of the Cd orbitals near the Fermi level ($E_\mathrm{F}$) is small compared with that of the other elements. Therefore, the overall behavior is qualitatively similar to the calculated results for the related compound~\prmgni~\cite{Wang2011-prmgni4-dos}.

To construct a low-energy effective model from the calculated DFT bands, we derive a 104-band tight-binding (TB) model using symmetry-protected maximally projected Wannier functions \cite{Eschrig2009-wannier,Koepernik2023-wannier}, comprising the Pr $4f,~5d$, Cd $5s,~5p$, and Ni $3d,~4s,~4p$ orbitals. As shown in Fig.~\ref{fig:band_DFT}(a), the resulting TB bands (blue dashed lines) accurately reproduce the DFT bands (black solid lines). The constructed TB Hamiltonian is subsequently used as input for the DMFT calculations in Sec.~\ref{subsec:prcdni4-DMFT}.

\subsection{DMFT calculations}
\label{subsec:prcdni4-DMFT}

The strong correlations among the $4f$ electrons are treated within DMFT~\cite{Georges1996,Kotliar2006}. The effective $j=5/2$ model of Eqs.~\eqref{eq:effjj-H}--\eqref{eq:effjj-int-matele} allows us to restrict the discussion of the $4f$ electrons to the $j=5/2$ subspace. The single-particle Green's function for 96 orbitals is given by
\begin{align}
    \label{eq:green}
      \hat{G}(\bm{k},i\omega_n)
      = \Big[ & (i\omega_n+\mu)\hat{I}-\hat{H}_{\mathrm{DFT}}(\bm{k})-\hat{H}^{(f^1)}_{\mathrm{CEF}} \notag \\
              & -\hat{\Sigma}_{\mathrm{loc}}(i\omega_n)-\varepsilon_f\hat{P}_f \Big]^{-1},
\end{align}
where $\hat{I}$ is the identity matrix. $\hat{H}_{\mathrm{DFT}}(\bm{k})$ is the TB Hamiltonian obtained from the DFT calculation, from which the CEF potential acting on the $4f$ electrons is removed. $\hat{H}^{(f^1)}_{\mathrm{CEF}}$ is the CEF potential with the parameters $w$ and $x$ [including the $1/\lambda$ correction as shown in Eq.~\eqref{eq:effjj-CEF}]. $\hat{\Sigma}_{\mathrm{loc}}(i\omega_n)$ is the local self-energy in the DMFT. $\varepsilon_f$ is the energy of the $4f$ level, and $\hat{P}_f$ is the projection operator onto the $4f$ orbitals. This term serves as the double-counting correction between the DFT and the DMFT.

Using the DFT estimates for the CEF potential does not reproduce the CEF levels in experiments. Therefore, we adjust the parameters $w$ and $x$ to reproduce the experimental CEF levels, as in the previous applications to Ce compounds~\cite{itokazu2025-cerh6ge4,Numa2026-cerh2as2}.
The explicit procedure for adjusting these parameters is presented in Appendix~\ref{app:adjusting-wx}.
We note that enforcing charge self-consistency between DFT and DMFT can improve the CEF levels and could allow us to avoid the manual adjustment~\cite{Delange2017,Pourovskii2020}.

Since the experimental properties of this compound~\cite{Kusanose2025-prcdni4} are well described by the localized $4f$-electron picture, we compute the self-energy $\hat{\Sigma}_{\mathrm{loc}}$ by exact diagonalization within the Hubbard-I approximation~\cite{Locht2016}. In performing this calculation, we need to adjust the Slater integrals of the Coulomb interaction, $F^0,\;F^2,\;F^4$, and $F^6$, the spin-orbit coupling constant $\lambda$, as well as the CEF parameters $w$ and $x$. First, the Coulomb interactions and $\varepsilon_f$ are chosen as follows. The $4f$ level is set to $\varepsilon_f=-8.0\;\mathrm{eV}$ to reproduce the photoemission spectroscopy results for elemental Pr, which show the $4f^2\rightarrow 4f^1$ excitation at $\Delta_-=3.3\;\mathrm{eV}$ and the $4f^2\rightarrow 4f^3$ excitation at $\Delta_+=2.1\;\mathrm{eV}$~\cite{Lang1981-PES}. The direct Coulomb interaction is taken as $U=5.7\;\mathrm{eV}$ and the Hund’s coupling as $J_{\mathrm{H}}=0.84\;\mathrm{eV}$. The corresponding Slater integrals are $F^0=5.7\;\mathrm{eV}$, $F^2=10\;\mathrm{eV}$, $F^4=6.7\;\mathrm{eV}$, and $F^6=4.9\;\mathrm{eV}$~\cite{Anisimov1997}.

\begin{figure}[t]
      \includegraphics[width=0.5\linewidth]{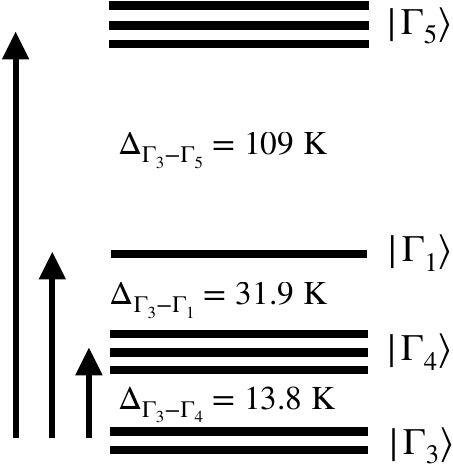}
      \caption{The CEF level scheme of the $J=4$ manifold of the $4f^2$ configuration obtained by fitting based on Sec.~\ref{sec:CEF}. $\Delta_{\Gamma_3-\Gamma_4},~\Delta_{\Gamma_3-\Gamma_1}$ and $\Delta_{\Gamma_3-\Gamma_5}$ denote the energies of the $\Gamma_4,~\Gamma_1$ and $\Gamma_5$ states, respectively, relative to the $\Gamma_3$ state.}
      \label{fig:CEF_DFT}
\end{figure}

Next, the spin-orbit coupling constant is taken as $\lambda=0.2\;\mathrm{eV}$, a typical value for $f$-electron compounds within the applicable range of the $j$--$j$ coupling description discussed in Sec.~\ref{sec:CEF}. Diagonalizing the effective $j=5/2$ Hamiltonian in Eqs.~\eqref{eq:effjj-H}--\eqref{eq:effjj-int-matele} with these parameters yields the CEF levels of the $4f^2$ ($J=4$) manifold shown in Fig.~\ref{fig:LLW}(d). Setting the CEF parameters to $w=4.23\,\mathrm{K}$ and $x=0.62$ so that $\Gamma_3$ becomes the ground state, we obtain the level scheme in Fig.~\ref{fig:CEF_DFT}. The resulting levels well reproduce the experimentally estimated CEF level scheme as shown in Fig.~\ref{fig:str}(b), in particular the $\Gamma_3$ ground state and the excitation energy to the first-excited $\Gamma_4$ state, $\Delta_{\Gamma_3-\Gamma_4}=13.8\,\mathrm{K}$.

The DFT+DMFT calculations were performed using open-source software DCore~\cite{Shinaoka2021-dcore} implemented on TRIQS~\cite{Parcollet2015-triqs} and TRIQS/DFTTools~\cite{Aichhorn2016-triqs-dft} libraries. We solved the effective impurity model by exact diagonalization using pomerol~\cite{Antipov2017-pomerol}.

\begin{figure}[t]
      \includegraphics[width=1.0\linewidth]{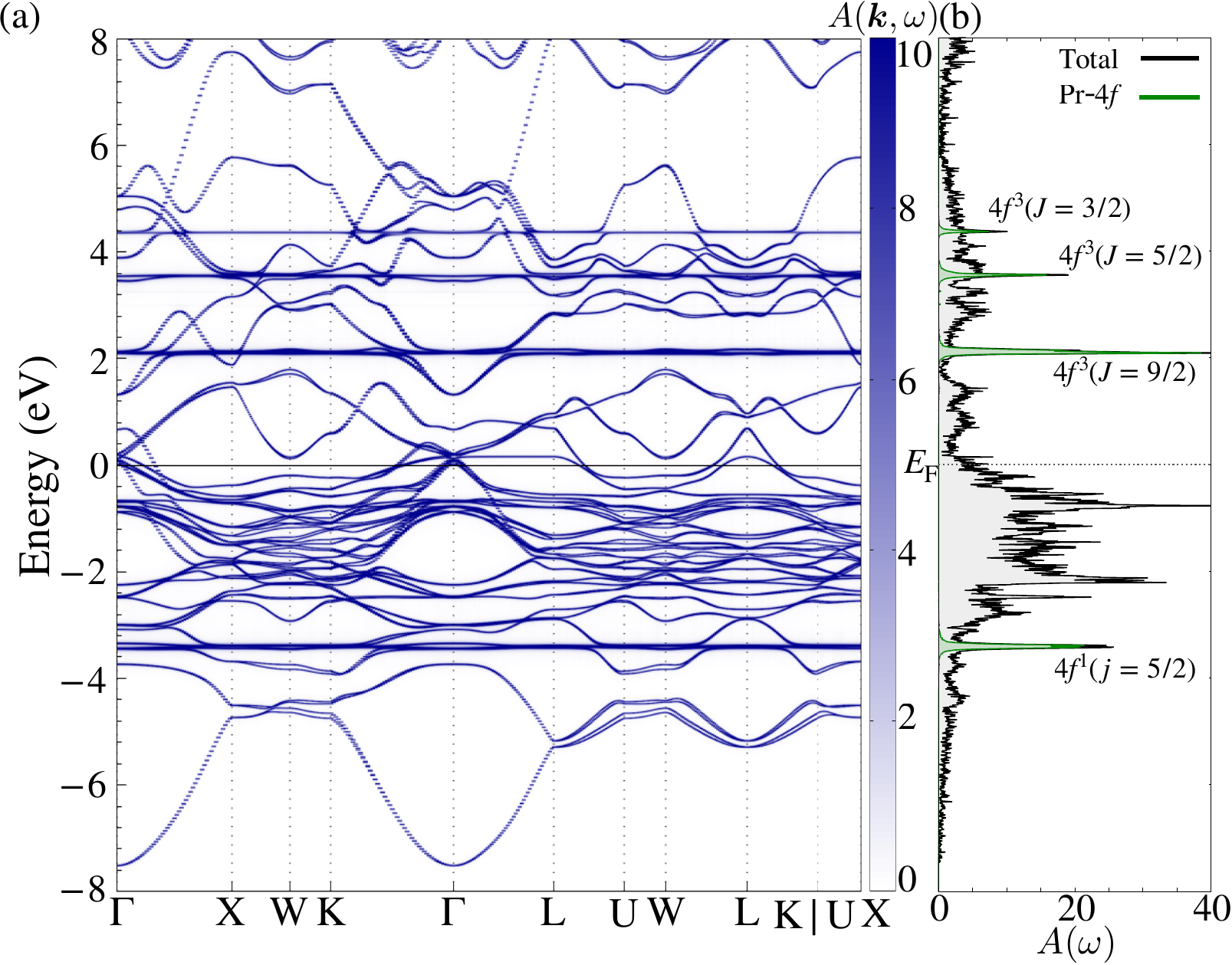}
      \caption{(a) The single-particle excitation spectrum $A(\bm{k},\omega)$ and (b) the $\bm{k}$-integrated spectrum $A(\omega)$ computed using the DFT+DMFT method.}
      \label{fig:band_DMFT}
\end{figure}

Figure~\ref{fig:band_DMFT} shows the single-particle excitation spectrum obtained from the DFT+DMFT calculation, where Fig.~\ref{fig:band_DMFT}(a) is the $\bm{k}$-resolved spectrum $A(\bm{k},\omega)$ and Fig.~\ref{fig:band_DMFT}(b) is the $\bm{k}$-integrated spectrum $A(\omega)$. The itinerant $4f$ bands near $E_{\mathrm{F}}$ in the DFT calculation in Fig.~\ref{fig:band_DFT}(a) are now located around $-\Delta_-=-3.3\;\mathrm{eV}$ and $+\Delta_+=2.1\;\mathrm{eV}$, reflecting the localized character of the $4f$ electrons.

\subsection{Multipole degrees of freedom}
\label{subsec:prcdni4-multipole}

We perform a symmetry analysis of possible fluctuation modes using multipoles, which enables a systematic characterization of symmetry-broken states~\cite{Kuramoto2009-review,Hayami2024-review}. We constructed 81 atomic multipoles of rank $0$--$8$ belonging to the $J=4$ multiplet using MultiPie~\cite{Kusunose2023-multipie}. 
Among them, here we focus on multipoles that are active within the CEF ground-state doublet $\Gamma_3$ and those appearing through mixing between $\Gamma_3$ and the first-excited triplet $\Gamma_4$ under the $T_d$ symmetry of the Pr site.

In group theory, the operators within the $\Gamma_3$ doublet are decomposed as follows:
\begin{align}
      \label{eq:G3-irrep}
      \Gamma_3 \otimes \Gamma_3 = A_{1}^{+} \oplus A_{2}^{-} \oplus E^{+},
\end{align}
where the superscript $+$ ($-$) indicates the time-reversal-even (odd) component, which appears by a symmetric (antisymmetric) product between two irreducible representations~\cite{inui2012group}. These operators are summarized in Table~\ref{tab:atomic_multipoles} (the second column). $A_1^+$ corresponds to the charge operator $Q_0$, and $A_2^-$ and $E^+$ correspond to the magnetic octupole $M_{xyz}$ and the electric quadrupoles $\{Q_{3z^2-r^2},\, Q_{x^2-y^2}\}$, respectively. It is important to note the absence of magnetic-dipole degrees of freedom.

\begin{table}[t]
      \caption{Atomic multipoles of $4f$ electrons. The column “Irrep” indicates the irreducible representation of the $T_d$ point group. The second and third columns show the four multipoles in the non-Kramers $\Gamma_3$ doublet system and the additional 12 multipoles that appear by mixing of the $\Gamma_3$ and the $\Gamma_4$ triplet, respectively. The notation follows Ref.~\cite{Hayami2024-review}.}
      \label{tab:atomic_multipoles}
      \begin{tabular}{llll}\hline
            Irrep & $\Gamma_3 \otimes \Gamma_3$ & $\Gamma_3 \otimes \Gamma_4$ \\
            \hline
            $A_{1}^{+}$ & $Q_{0}$ & \\
            $A_{2}^{+}$ & & \\
            $E^{+}$ & $\{Q_{3z^2-r^2},~Q_{x^2-y^2}\}$ & \\
            $T_{1}^{+}$ & & $\{Q_{yz(y^2-z^2)},~Q_{zx(z^2-x^2)},~Q_{xy(x^2-y^2)}\}$ \\
            $T_2^{+}$   & & $\{Q_{yz},~Q_{zx},~Q_{xy}\}$ \\
            \hline
            $A_{1}^{-}$ & & \\
            $A_{2}^{-}$ & $M_{xyz}$ & \\
            $E^{-}$ & & \\
            $T_{1}^{-}$ & & $\{M_{x},~M_{y},~M_{z}\}$ \\
            $T_{2}^{-}$ & & $\{M_{x(y^2-z^2)},~M_{y(z^2-x^2)},~M_{z(x^2-y^2)}\}$ \\
            \hline
      \end{tabular}
\end{table}

We next consider multipoles that appear by mixing of the ground-state doublet $\Gamma_3$ and the first-excited triplet $\Gamma_4$.
The symmetries of the multipole operators are represented as
\begin{align}
      \label{eq:G3G4-irrep}
      \Gamma_3 \otimes \Gamma_4 = T_{1}^{\pm} \oplus T_{2}^{\pm},
\end{align}
where the superscript $\pm$ indicates that both electric and magnetic operators exist.
The third column of Table~\ref{tab:atomic_multipoles} summarizes these operators. The operators in $T_1^{+}$ are higher-order electric hexadecapoles, and those in $T_2^{+}$ are electric quadrupoles. The operators in $T_1^{-}$ are magnetic dipoles, and those in $T_2^{-}$ are magnetic octupoles.
Therefore, magnetic-dipole degrees of freedom become active through the mixing between $\Gamma_3$ and $\Gamma_4$, which is expected at temperatures comparable or higher than the CEF splitting.

\subsection{Multipolar susceptibilities}
\label{subsec:prcdni4-susceptibility}

In DMFT, $\chi(\bm{q})$ can be computed by solving the Bethe-Salpeter equation for the two-particle Green's function~\cite{Georges1996}. In this paper, we use the following strong-coupling limit (SCL) formula of Refs.~\cite{Otsuki2019-SCL,Otsuki2024-ceb6}, which is valid for well-localized systems such as {\prcdni}:
\begin{align}
      \label{eq:chi_q_matrix}
      \hat{\chi}(\bm{q}) = [\hat{\chi}_\mathrm{loc}^{-1} - \hat{I}(\bm{q})]^{-1}.
\end{align}
Here, the hat denotes a matrix with respect to the combined indices $\alpha_1\alpha_2$ and $\alpha_3\alpha_4$ of $\chi$ in Eq.~\eqref{eq:f1chi-def}. $\hat{\chi}_{\mathrm{loc}}$ is the local susceptibility evaluated from the effective impurity problem in DMFT. $\hat{I}(\bm{q})$ describes the effective interaction, which is calculated using the excitation energies $\Delta_-$ and $\Delta_+$ extracted from $A(\bm{k},\omega)$ (see Sec.~\ref{subsec:prcdni4-DMFT}). 
The susceptibility calculations were performed using the open-source package ChiQ~\cite{ChiQ}.

To discuss multipolar fluctuations of the $4f^2$ electrons in the present system, we transform the quantities represented in the single-particle basis ($\hat{\chi}$, $\hat{\chi}_{\mathrm{loc}}$, and $\hat{I}$) into those in the many-particle basis ($\hat{\mathcal{X}}$, $\hat{\mathcal{X}}_{\mathrm{loc}}$, and $\hat{\mathcal{I}}$), based on the procedure described in Sec.~\ref{sec:fn-fluctuation}.
As discussed in Sec.~\ref{sec:fn-fluctuation} and Appendix~\ref{app:over-under}, the mapping from the $j=5/2$ ($4f^1$) basis to the $J=4$ ($4f^2$) basis is an underdetermined problem.
Thus, the many-particle quantities such as $\mathcal{X}_{M_1M_2,M_3M_4}$ computed in Eq.~\eqref{eq:f1-to-f2chi} correspond to the minimum-$L_2$-norm solution.

We employ two ways for the extraction of the multipolar components of $\mathcal{X}(\bm{q})$, $\mathcal{X}_{\mathrm{loc}}$, and $\mathcal{I}(\bm{q})$. First, we diagonalize the matrix as follows:
\begin{align}
    \label{eq:chi-eig}
    \mathcal{X}^{(\xi)}(\bm{q})=\sum_{M_1M_2M_3M_4}u^{(\xi)}_{M_1M_2}(\bm{q})\mathcal{X}_{M_1M_2,M_3M_4}(\bm{q})u^{(\xi)}_{M_3M_4}(\bm{q})^{\ast},
\end{align}
where the eigenmodes are labeled with the superscript $\xi$. The eigenvalues $\mathcal{X}^{(\xi)}(\bm{q})$ represent the magnitude of the fluctuations and the eigenvectors $u^{(\xi)}_{M_1M_2}$ represent the corresponding magnetic and orbital configurations. The eigenvectors $u^{(\xi)}$ are projected onto the multipole basis $\hat{Z}_\gamma$, yielding the expansion coefficients $c^{(\xi)}_\gamma=\sum_{M_1M_2}(\hat{Z}_\gamma)_{M_1M_2}^{\ast}u^{(\xi)}_{M_1M_2}$. The character of each eigenmode is then classified according to the irreducible representation that gives the maximum weight $|c^{(\xi)}_\gamma|^2$~\cite{Numa2026-cerh2as2}. Second, we take linear combinations of the matrix using the multipole basis $\hat{Z}_\gamma$:
\begin{align}
\label{eq:chi-lc}
\mathcal{X}_{\gamma}(\bm{q})=\sum_{M_1M_2M_3M_4}
(\hat{Z}_{\gamma})_{M_1M_2}\mathcal{X}_{M_1M_2,M_3M_4}(\bm{q})(\hat{Z}_{\gamma})^{\ast}_{M_3M_4},
\end{align}
where $\mathcal{X}_{\gamma}(\bm{q})$ [and similarly $\mathcal{I}_{\gamma}(\bm{q})$] represents the $\gamma$ component of the multipolar susceptibility.

\begin{figure}[t]
      \includegraphics[width=1.0\linewidth]{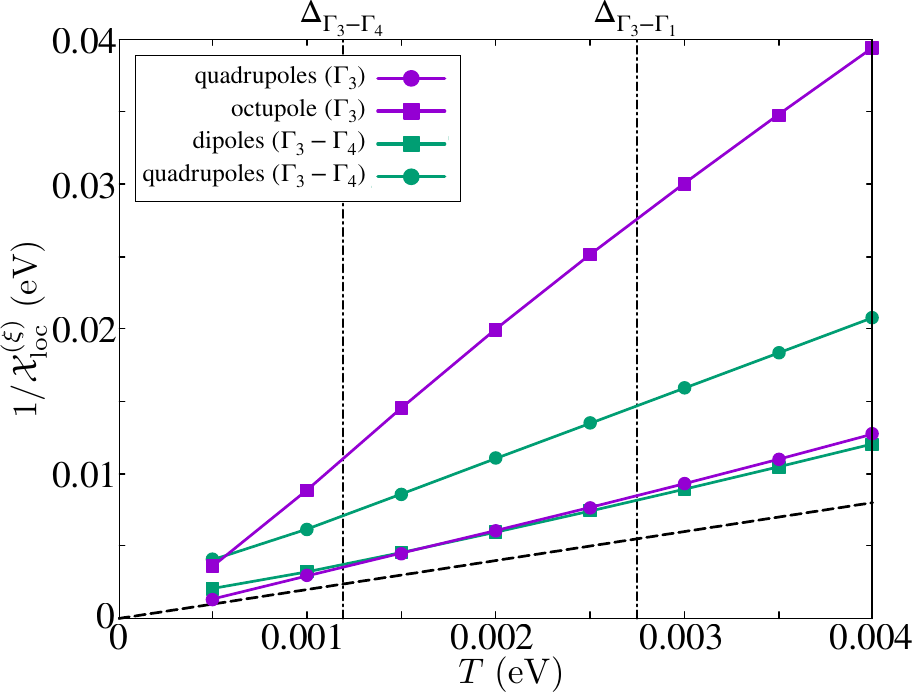}
      \caption{Temperature dependence of the local susceptibilities $\mathcal{X}^{(\xi)}_{\mathrm{loc}}$. The dashed line shows the Curie law at low $T$. The vertical dash-dotted lines indicate the energies of the CEF splitting between the ground state $\Gamma_3$ and the excited states $\Gamma_4$ and $\Gamma_1$ in Fig.~\ref{fig:CEF_DFT}.}
      \label{fig:t-chiinv}
\end{figure}

Before discussing the $\bm{q}$ dependence, we examine the temperature dependence of the local susceptibility $\mathcal{X}^{(\xi)}_{\mathrm{loc}}$ as shown in Fig.~\ref{fig:t-chiinv}. At temperatures well above $\Delta_{\Gamma_3-\Gamma_4}=0.0012$~eV, the magnetic dipolar fluctuations ($T_1^-$) arising from the product $\Gamma_3\otimes\Gamma_4$ are the largest. As the temperature decreases, the dipolar fluctuations are suppressed and the electric quadrupolar fluctuations ($E^+$) originating from the $\Gamma_3$ doublet develop. For $T \lesssim \Delta_{\Gamma_3-\Gamma_4}$, the quadrupolar fluctuations eventually overtake the dipolar fluctuations and become dominant. Specifically, the quadrupolar susceptibility approaches the Curie law, 
\begin{align}
    \label{eq:curie}
    \mathcal{X}^{(\xi)}_{\mathrm{loc}}\approx\frac{C}{T},
\end{align}
with the Curie constant $C = 1/2$.
These behaviors become clearer for a larger CEF splitting, which is presented in Appendix~\ref{app:t-chiloc-10w}.
The appendix also provides a comparison between $\mathcal{X}_{\mathrm{loc}}$ in the $4f^2$ basis and $\chi_{\mathrm{loc}}$ in the $4f^1$ basis.

\begin{figure}[t]
      \includegraphics[width=1.0\linewidth]{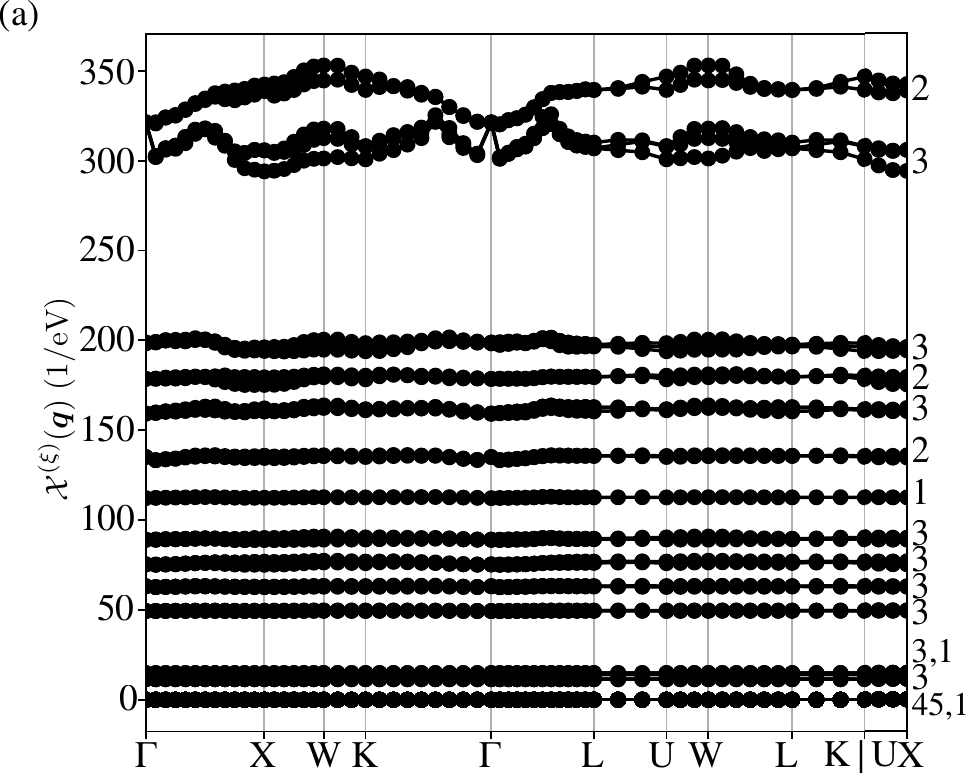}
      \includegraphics[width=1.0\linewidth]{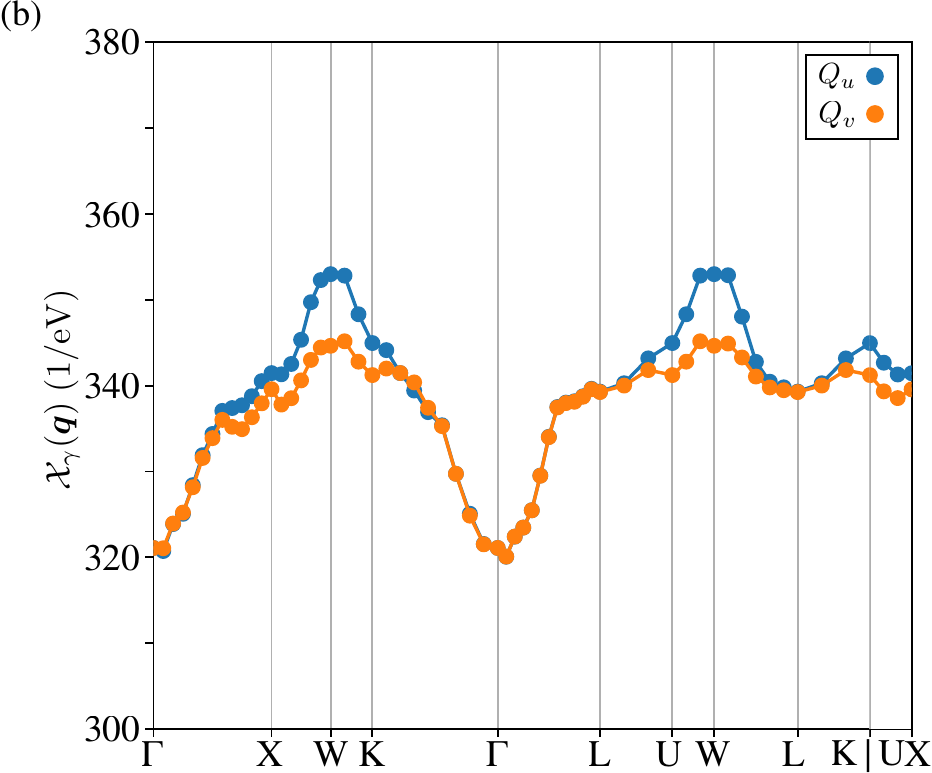}
      \caption{(a) Eigenvalues $\mathcal{X}^{(\xi)}(\bm{q})$ of the susceptibility at $T=1\;\mathrm{meV}$. The numbers on the right indicate the number of modes in each cluster of $\mathcal{X}^{(\xi)}(\bm{q})$.
        (b) The $\bm{q}$ dependence of the quadrupolar susceptibility $\mathcal{X}_{\gamma}(\bm{q})$ within the $\Gamma_3$ degrees of freedom. The $\bm{q}$-path is defined in Fig.~\ref{fig:band_DFT}(c).}
      \label{fig:f2chiq-all}
\end{figure}

Next, we discuss the $\bm{q}$ dependence of the eigenvalues of the transformed $\mathcal{X}$. Figure~\ref{fig:f2chiq-all}(a) shows the $\bm{q}$ dependence of all eigenvalues $\mathcal{X}^{(\xi)}(\bm{q})$ at $T=1\;\mathrm{meV}$. There are $9^2=81$ modes in total, corresponding to the atomic degrees of freedom within $J=4$. The overall structure can be understood in terms of the CEF states. The largest two fluctuation modes are the electric quadrupolar fluctuations within the $\Gamma_3$ ground-state doublet. The second group with three modes corresponds to magnetic dipolar fluctuations arising from mixing between the ground state and the first-excited $\Gamma_4$ triplet. The single mode at $\mathcal{X}^{(\xi)}(\bm{q})=113~(1/\mathrm{eV})$ corresponds to a magnetic octupolar fluctuation within $\Gamma_3$. The lowest (almost zero) single-mode fluctuation is the charge fluctuation. The other $45$ ($=81-36$) zero modes result from the lack of information in the mapping from $4f^1(j=5/2)$ to $4f^2(J=4)$, and correspond to higher-rank multipoles (rank 6 and above).

We focus on the two largest modes.
Figure~\ref{fig:f2chiq-all}(b) shows the $\bm{q}$ dependence of the quadrupolar susceptibilities $\mathcal{X}_{\gamma}(\bm{q})$ with $\gamma=\{u,v\}$.
Here, we evaluated $\mathcal{X}_{\gamma}(\bm{q})$ by taking the linear combination in Eq.~\eqref{eq:chi-lc} using the eigenvectors of the two largest modes of $\mathcal{X}^{(\xi)}(\bm{q})$ at $\bm{q}=\QW\equiv(2\pi,\pi,0)$ as the basis $\hat{Z}_{\gamma}$.
From the overlap between $\hat{Z}_{\gamma}$ and the MultiPie basis, we identified each mode as $Q_u$ or $Q_v$.
The resulting dispersion of $\mathcal{X}_{\gamma}(\bm{q})$ well reproduces that of $\mathcal{X}^{(\xi)}(\bm{q})$, confirming the validity of the chosen basis.
As seen in Fig.~\ref{fig:f2chiq-all}(b), the leading instability occurs in the $Q_u$ channel at $\bm{q}=\QW$, indicating a $Q_u$-type antiferroquadrupolar order with $\bm{q}= \QW$.

\subsection{Quadrupolar interaction}
\label{subsec:prcdni4-interaction}

We derive the effective interactions $\mathcal{I}(\bm{q})$ between multipoles. This quantity allows us to access lower temperatures to estimate the transition temperature.
Furthermore, its real-space form $\mathcal{I}(\bm{r})$ can be used to construct the effective Hamiltonian in Sec.~\ref{subsec:prcdni4-effH}.

\begin{figure}[t]
      \includegraphics[width=1.0\linewidth]{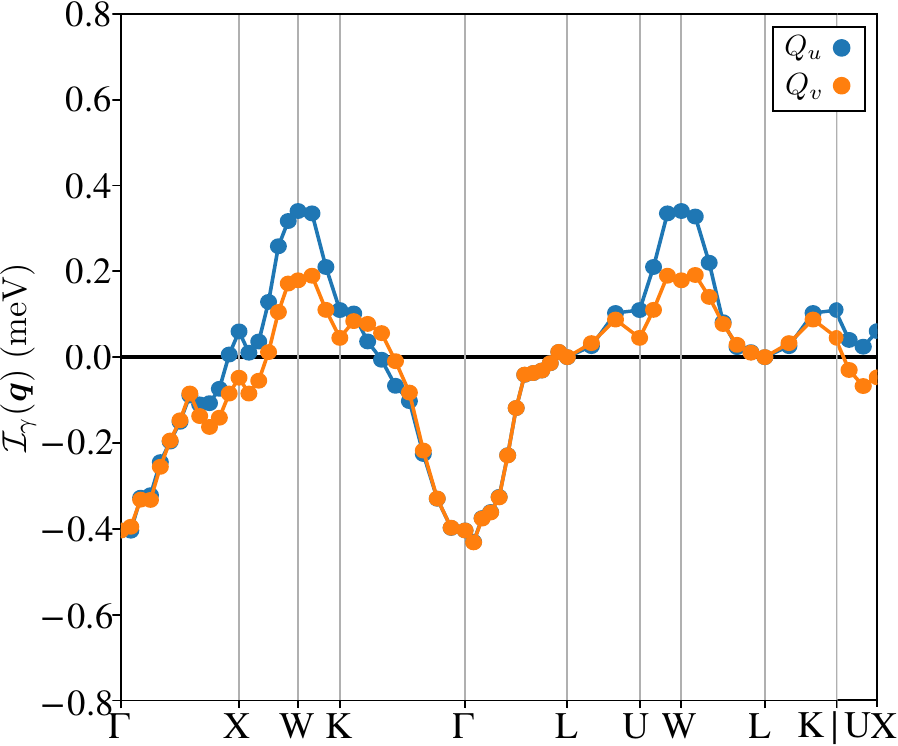}
      \caption{The momentum-dependent effective quadrupolar interaction $\mathcal{I}_{\gamma}(\bm{q})$ within the $\Gamma_3$ degrees of freedom.}
      \label{fig:Iq_Wpoint}
\end{figure}

Figure~\ref{fig:Iq_Wpoint} shows the quadrupolar interactions $\mathcal{I}_{\gamma}(\bm{q})$ with $\gamma=\{u,v\}$, evaluated by the linear combination in Eq.~\eqref{eq:chi-lc}.
The $\bm{q}$ dependence closely follows that of $\mathcal{X}_{\gamma}(\bm{q})$ in Fig.~\ref{fig:f2chiq-all}(b), and the interactions are enhanced near the W point.
For the leading fluctuation mode, that is, the electric quadrupole $Q_u$ at $\bm{q}=\QW$, we obtain $\mathcal{I}_{u}(\QW)=0.341\;\mathrm{meV}$.
This value together with the Curie law for $\mathcal{X}^{(\xi)}_{\mathrm{loc}}$ in Eq.~\eqref{eq:curie} can be used to extrapolate $\mathcal{X}_{\gamma}(\bm{q})$ to lower temperatures and find its divergence. The transition temperature is thus estimated to be $2.0\;\mathrm{K}$, which overestimates the experimental value $T_{\mathrm O}\approx1.0\;\mathrm{K}$ by a factor of two but reproduces the correct order of magnitude.

We next derive real-space interactions. We compute $I(\bm{q})$ over the entire Brillouin zone and perform a Fourier transform to obtain its real-space counterpart $I(\bm{r})$. The intersite interaction $I_{m_1m_2,m_3m_4}(\bm{r})$, written in the single-particle basis, is then transformed to $\mathcal{I}_{M_1M_2,M_3M_4}(\bm{r})$ represented in the many-particle basis using the transformation as in Eq.~\eqref{eq:f1-to-f2chi}. The quadrupolar components of the interaction $\mathcal{I}_{\gamma}(\bm{r}_i-\bm{r}_j)$ with $\gamma=\{u,v\}$ define the effective Heisenberg-type Hamiltonian as
\begin{align}
      \label{eq:Heisenberg}
      \mathcal{H}=-\frac{1}{2}\sum_{ij,\gamma=\{u,v\}}Q_{\gamma}(\bm{r}_i)\mathcal{I}_{\gamma}(\bm{r}_i-\bm{r}_j)Q_{\gamma}(\bm{r}_j).
\end{align}
A positive (negative) value of $\mathcal{I}_{\gamma}(\bm{r})$ indicates ferroic (antiferroic) coupling.

\begin{figure}[t]
      \includegraphics[width=0.6\linewidth]{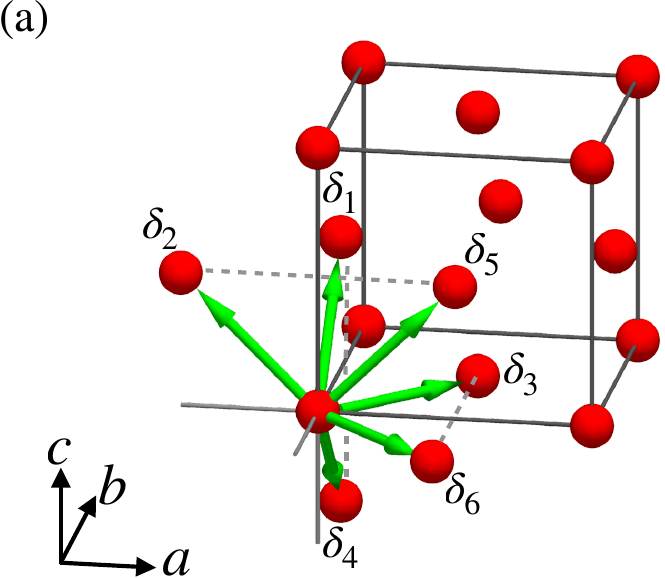}
      \includegraphics[width=1.0\linewidth]{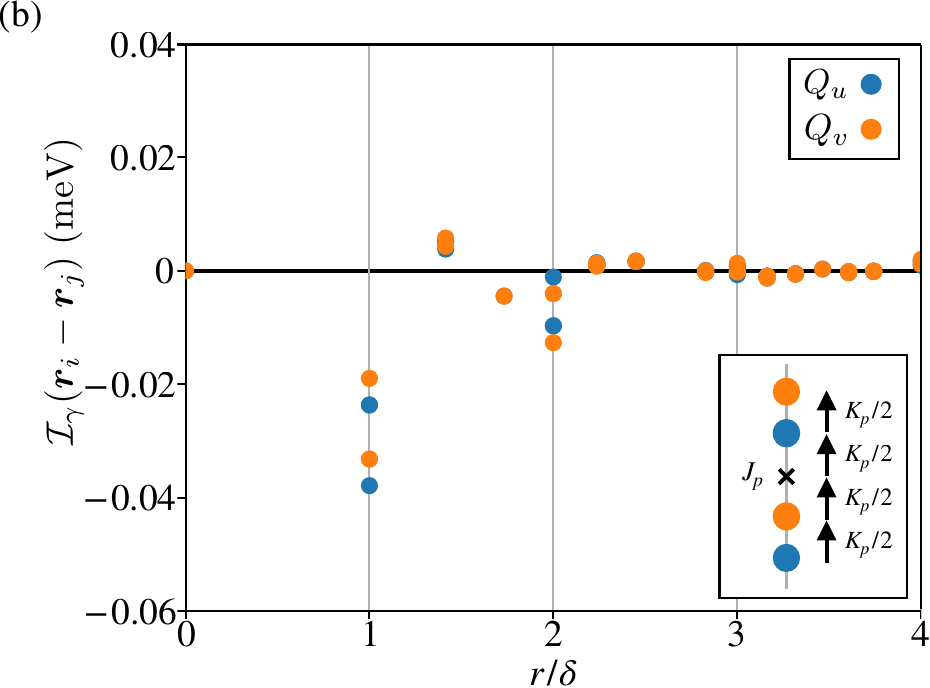}
      \caption{(a) Schematic picture of the nearest-neighbor bond vectors $\bm{\delta}_n$ on the fcc lattice of Pr ions.
            (b) The effective interactions $\mathcal{I}_{\gamma}(\bm{r}_i-\bm{r}_j)$ between quadrupole moments as a function of the distance $r \equiv |\bm{r}_i-\bm{r}_j|$. The distance is normalized by $\delta \equiv |\bm{\delta}_n|$. The inset shows the eigenvalues of $\hat{J}_p$ in Eq.~\eqref{eq:HQ-Jn} and their relation to the parameters $J_p$ and $K_p$.}
      \label{fig:Ir_Wpoint}
\end{figure}

Figure~\ref{fig:Ir_Wpoint}(a) illustrates the six nearest-neighbor (NN) bond vectors $\bm{\delta}_n$ ($n=1, \cdots, 6$) connecting Pr sites in the fcc lattice. In units of the lattice constant $a$, these vectors are defined as $\bm{\delta}_{1,4}=(0,1/2,\pm1/2)$, $\bm{\delta}_{2,5}=(\mp1/2,0,1/2)$, and $\bm{\delta}_{3,6}=(1/2,\pm1/2,0)$. Here, the first and second indices of $\bm{\delta}$ correspond to the upper and lower signs on the right-hand side, respectively.

Figure~\ref{fig:Ir_Wpoint}(b) shows the distance dependence of the intersite interaction $\mathcal{I}_{\gamma}(\bm{r}_i-\bm{r}_j)$, where the distance is measured in units of the NN bond length $|\bm{\delta}_n| \equiv \delta$. The dominant contribution arises from the antiferroquadrupolar $Q_u$ component between NN fcc sites, corresponding to $|\bm{r}_i-\bm{r}_j|=\delta$, taking a value of $-0.038\;\mathrm{meV}\approx-440$ mK. The next largest contribution is the quadrupolar interaction between fourth-nearest-neighbor (fourth-NN) sites, corresponding to $|\bm{r}_i-\bm{r}_j|=2\delta$. In contrast to the NN case, this interaction is dominated by the $Q_v$ component, with a value of $-0.013\;\mathrm{meV}\approx-147$ mK. Our calculations reveal that both the NN and fourth-NN interactions exhibit spatial anisotropy. Specifically, $\mathcal{I}_{\gamma}$ has the same strength along the in-plane directions ($\bm{\delta}_{3}$ and $\bm{\delta}_{6}$) and along the out-of-plane directions ($\bm{\delta}_{1}$, $\bm{\delta}_{2}$, $\bm{\delta}_{4}$, and $\bm{\delta}_{5}$), as summarized in Table~\ref{tab:Ir_values}. It will be demonstrated in Sec.~\ref{subsec:prcdni4-effH} that the presence of the fourth-NN interaction is essential for stabilizing the quadrupolar order at the wave vector $\QW$ found in the susceptibility calculation.

\begin{table}[t]
      \caption{Quadrupolar intersite interactions $\mathcal{I}_{\gamma}$ with $\gamma=\{u,v\}$ along the fcc bonds $\bm{\delta}_n$ (nearest neighbor) and $2\bm{\delta}_n$ (fourth-nearest neighbor). The bond vectors are defined in Fig.~\ref{fig:Ir_Wpoint}(a).}
      \label{tab:Ir_values}
      \begin{tabular}{lll}\hline
            bonds & $\mathcal{I}_u$~(meV) & $\mathcal{I}_v$~(meV) \\ \hline
            $\bm{\delta}_1$, $\bm{\delta}_2$, $\bm{\delta}_4$, $\bm{\delta}_5$ & $-0.024$ & $-0.033$ \\
            $\bm{\delta}_3$, $\bm{\delta}_6$ & $-0.038$ & $-0.019$ \\
            $2\bm{\delta}_1$, $2\bm{\delta}_2$, $2\bm{\delta}_4$, $2\bm{\delta}_5$ & $-0.0097$ & $-0.0040$ \\
            $2\bm{\delta}_3$, $2\bm{\delta}_6$ & $-0.0011$ & $-0.013$ \\
            \hline
      \end{tabular}
\end{table}

\subsection{Discussion}
\label{subsec:prcdni4-effH}

Based on the numerical results so far, we construct an effective Hamiltonian for the quadrupolar degrees of freedom in the $\Gamma_3$ doublet.
Symmetry analysis shows that the NN quadrupolar interaction on the fcc lattice is characterized by two parameters~\cite{Tsunetsugu2021,Hattori2023}.
Incorporating the fourth-NN interactions as well, the effective Hamiltonian is given by
\begin{align}
      \label{eq:HQ}
      H_Q=-\sum_{p=1,4}\sum_{\langle\bm{r},\bm{r}'\rangle_p}\bm{Q}(\bm{r})\cdot \hat{J}_p(\bm{r}-\bm{r}')\bm{Q}(\bm{r}'), \\
      \label{eq:HQ-Jn}
      \hat{J}_p(\bm{r}-\bm{r}')=J_p\hat{\sigma}_0+K_p\hat{g}(\bm{r}-\bm{r}'),
\end{align}
where $\bm{Q}\equiv(Q_u,Q_v)^{\mathrm{T}}$, and the summation with $\langle\bm{r},\bm{r}'\rangle_p$ runs over pairs of NN ($p=1$) and fourth-NN ($p=4$) sites. The parameters $J_p$ and $K_p$ denote the isotropic and anisotropic interactions. Note that, to be consistent with our definition of the susceptibility in Eq.~(\ref{eq:chi_q_matrix}), we define the interaction parameters with signs opposite to those in Refs.~\cite{Tsunetsugu2021,Hattori2023}. 
$\hat{\sigma}_0$ is the $2\times2$ identity matrix. $\hat{g}(\bm{r}-\bm{r}')$ describes the bond-dependent anisotropy of the quadrupolar interaction, and is defined by
\begin{align}
      \label{eq:ani-factor}
      \hat{g}(\bm{r}-\bm{r}')=\mathrm{cos}\zeta(\bm{r}-\bm{r}')\hat{\sigma}_3-\mathrm{sin}\zeta(\bm{r}-\bm{r}')\hat{\sigma}_1,
\end{align}
where $\hat{\sigma}_1$ and $\hat{\sigma}_3$ are Pauli matrices. The angle $\zeta$ takes the specific values $\zeta(\bm{\delta}_{1})=\zeta(\bm{\delta}_{4})=-\pi/3$, $\zeta(\bm{\delta}_{3})=\zeta(\bm{\delta}_{6})=\pi$, and $\zeta(\bm{\delta}_{2})=\zeta(\bm{\delta}_{5})=\pi/3$ for the NN bonds. Since the fourth-NN bonds lie along the same directions, $\zeta(2\bm{\delta}_n)=\zeta(\bm{\delta}_n)$ holds.

The interaction parameters $J_1, K_1, J_4$, and $K_4$ are extracted from the numerical results as follows. 
The eigenvalues of the matrix $\hat{J}_p$ in Eq.~\eqref{eq:HQ-Jn} are distributed as shown in the inset of Fig.~\ref{fig:Ir_Wpoint}(b). $J_p$ determines the center of the distribution and $K_p$ yields its spread with equal spacing.
The NN and fourth-NN interactions $\mathcal{I}_{\gamma}$ shown in Fig.~\ref{fig:Ir_Wpoint}(b) and listed in Table~\ref{tab:Ir_values} coincide with this distribution within an accuracy of $1\;\mathrm{mK}$.
We thus obtain $J_1=-330\;\mathrm{mK}$, $K_1=109\;\mathrm{mK}$, $J_4=-79.8\;\mathrm{mK}$, and $K_4=-68.4\;\mathrm{mK}$.

\begin{figure}[t]
      \includegraphics[width=1.0\linewidth]{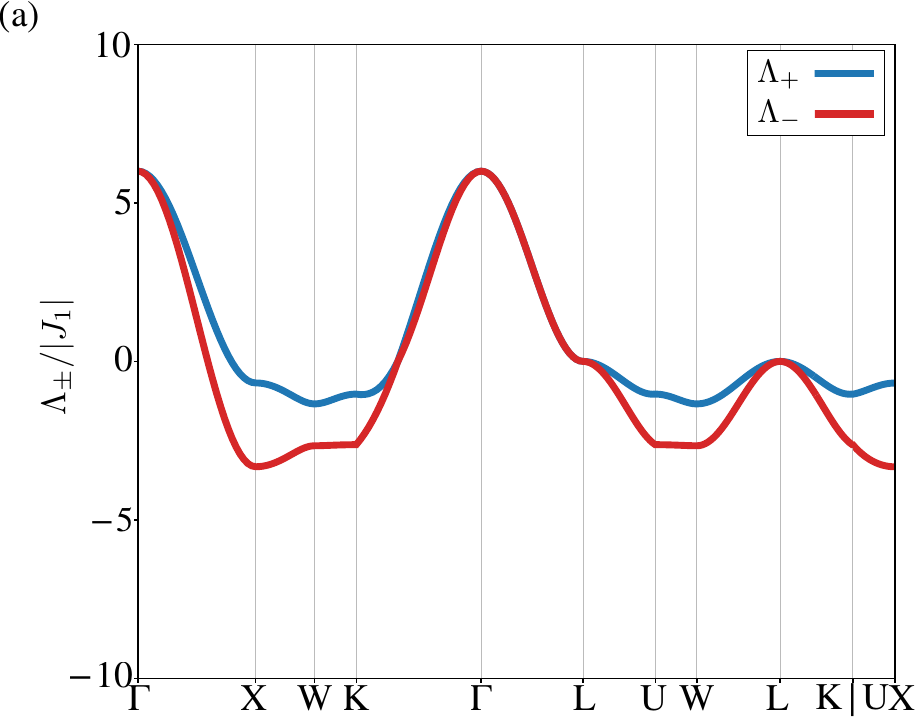}
      \includegraphics[width=1.0\linewidth]{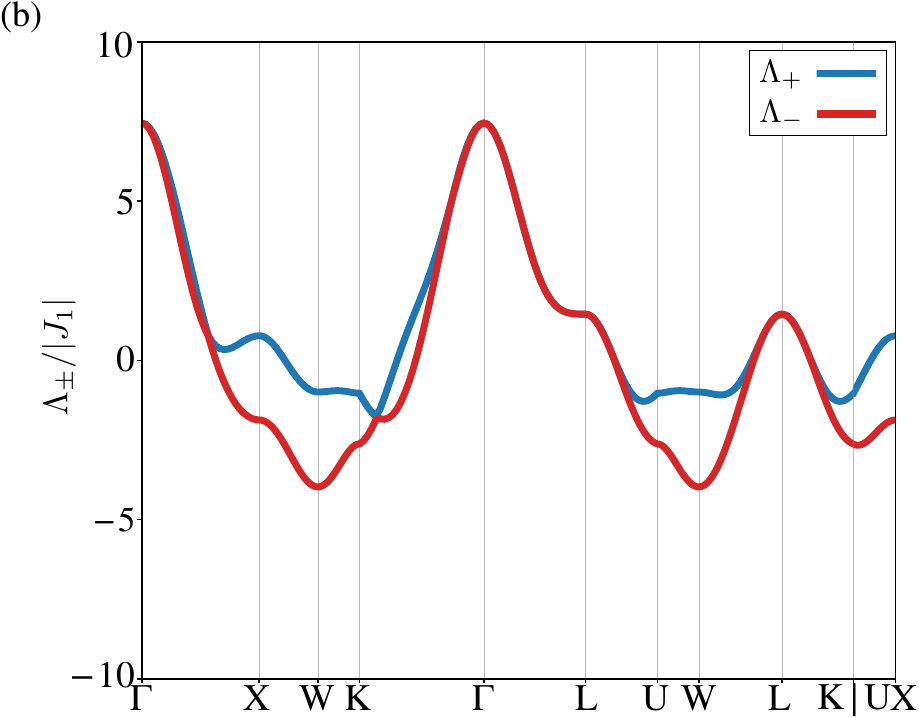}
      \caption{Eigenvalues of the effective Hamiltonian $H_Q$ in Eq.~\eqref{eq:HQk} with (a) only the NN interactions $J_1$ and $K_1$ and (b) additional fourth-NN interactions $J_4$ and $K_4$.}
      \label{fig:lambda}
\end{figure}

To investigate the classical ground state of $H_{Q}$, we perform a Fourier transform using $\bm{Q}_{\bm{q}} = \frac{1}{\sqrt{N}} \sum_{\bm{r}} \bm{Q}(\bm{r}) e^{-i\bm{q}\cdot\bm{r}}$, where $N$ is the total number of $\bm{q}$-points in the Brillouin zone. In momentum space, $H_Q$ is written as
\begin{align}
      \label{eq:HQk}
      H_Q=-\sum_{\bm{q}}\bm{Q}_{\bm{q}}\cdot \hat{J}(\bm{q})\bm{Q}_{-\bm{q}}.
\end{align}
By diagonalizing the interaction matrix $\hat{J}(\bm{q})$, we obtain the eigenenergies $\Lambda_{\pm}(\bm{q})$ of $H_{Q}$.
Figure~\ref{fig:lambda}(a) shows the $\bm{q}$-dependence of $\Lambda_{\pm}(\bm{q})$ calculated using the interaction parameters $J_1$ and $K_1$ obtained above with $J_4=K_4=0$, corresponding to the Hamiltonian with only the NN interactions. The eigenenergy is lowest at the X point, which is consistent with previous studies~\cite{Tsunetsugu2021,Hattori2023}. 
In contrast, as shown in Fig.~\ref{fig:lambda}(b), the inclusion of the fourth-NN interactions, $J_4$ and $K_4$, reproduces the dispersion of $\mathcal{I}_{\gamma}(\bm{q})$ in Fig.~\ref{fig:Iq_Wpoint}, with the energy minimum shifting to the W point.
Therefore, we conclude that the fourth-NN as well as NN interactions are relevant for stabilizing the quadrupolar ordering at the W point.

\begin{figure}[t]
      \includegraphics[width=1.0\linewidth]{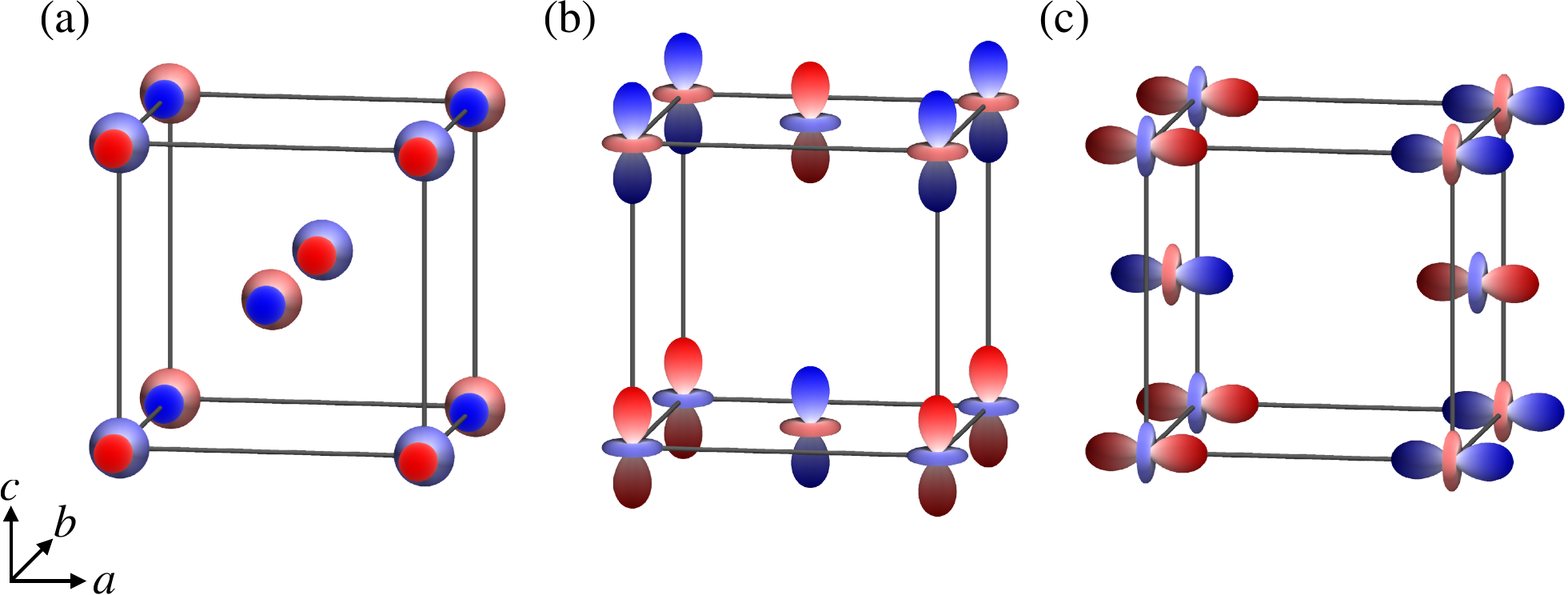}
      \caption{Stable quadrupolar order obtained from the classical eigenstates of the effective Hamiltonian in Eq.~\eqref{eq:HQk} for $K_1>0$ and $K_4<K_1/2$: (a) $Q_{3y^2-r^2}$ with $\bm{q}=(2\pi,~\pi,~0)$, (b) $Q_{3z^2-r^2}$ with $\bm{q}=(2\pi,~0,~\pi)$, and (c) $Q_{3x^2-r^2}$ with $\bm{q}=(\pi,~2\pi,~0)$.}
      \label{fig:Qu_singleq}
\end{figure}

\begin{figure}[t]
      \includegraphics[width=1.0\linewidth]{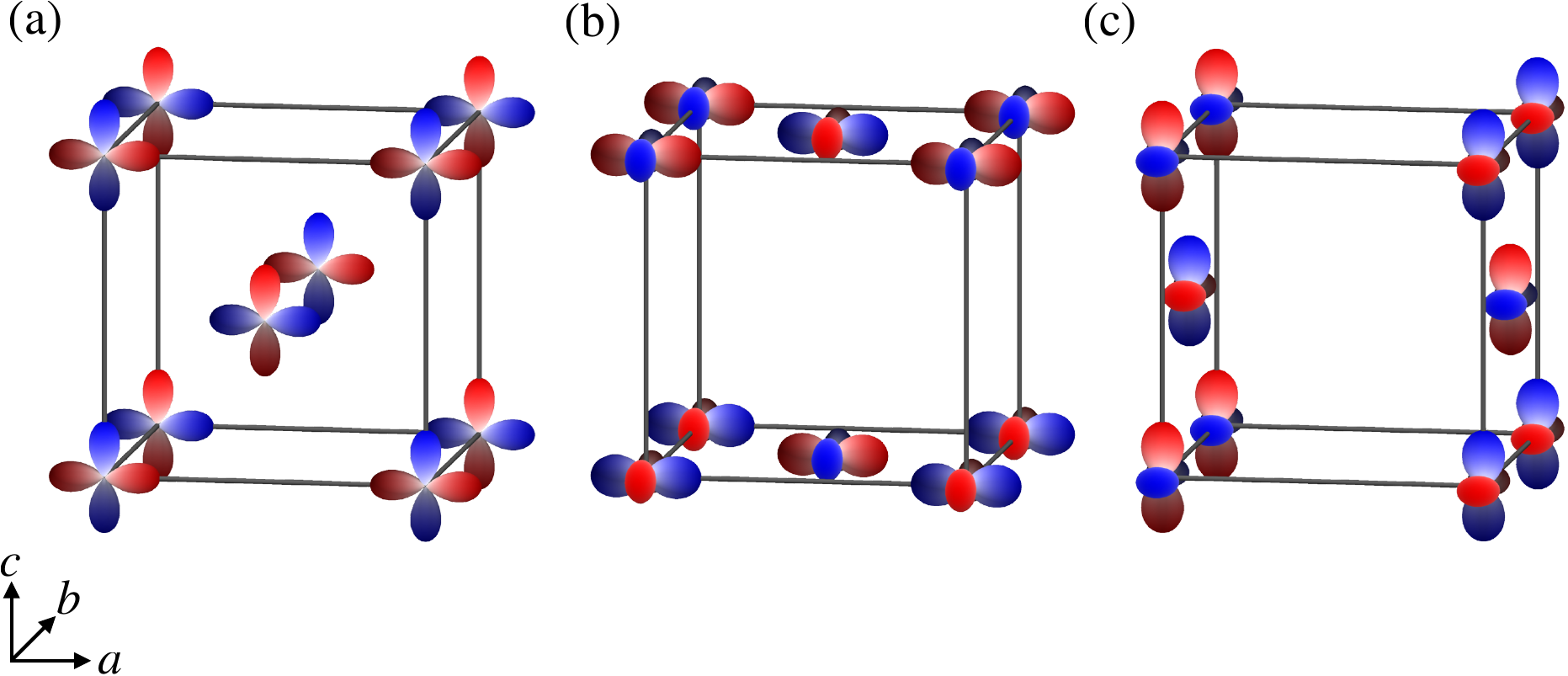}
      \caption{Stable quadrupolar order obtained from the classical eigenstates of the effective Hamiltonian in Eq.~\eqref{eq:HQk} for $K_1<0$ and $K_4>K_1/2$: (a) $Q_{z^2-x^2}$ with $\bm{q}=(2\pi,~\pi,~0)$, (b) $Q_{x^2-y^2}$ with $\bm{q}=(2\pi,~0,~\pi)$, and (c) $Q_{y^2-z^2}$ with $\bm{q}=(\pi,~2\pi,~0)$.}
      \label{fig:Qv_singleq}
\end{figure}

The real-space configuration of the quadrupolar moments for the W-point order can be determined from the eigenvector of $H_Q$. 
The $Q_u$-type quadrupoles are stabilized for the estimated parameter set with $K_1 > 0$ and $K_4 < 0$.
There are 24 equivalent W points in the Brillouin zone, which are reduced to three independent configurations, as shown in Fig.~\ref{fig:Qu_singleq}. For example, the eight wave vectors, $(2\pi,\pm\pi,0)$ and $(0,\;\pm\pi,2\pi)$ and their inverses, yield the configuration in Fig.~\ref{fig:Qu_singleq}(a). The $(3y^2-r^2)$-type quadrupoles are aligned antiferroically in the $a$--$c$ plane. The $y=0$ and $y=1$ layers have opposite signs, leaving Pr sites at $y=1/2$ disordered.
The configurations in Figs.~\ref{fig:Qu_singleq}(b) and \ref{fig:Qu_singleq}(c) are obtained by $90^{\circ}$ rotations of Fig.~\ref{fig:Qu_singleq}(a).
On the other hand, the $Q_v$-type quadrupolar state becomes the ground state in the opposite case, $K_1 < 0$ and $K_4 > 0$, as illustrated in Fig.~\ref{fig:Qv_singleq}.
The precise boundary between the configurations in Figs.~\ref{fig:Qu_singleq} and \ref{fig:Qv_singleq} is given by $K_1/2=K_4$.

We now consider the phase transition in {\prcdni} at $T_{\mathrm{O}}=1.0~\mathrm{K}$ based on our DFT+DMFT results.
As shown in Figs.~\ref{fig:Qu_singleq} and~\ref{fig:Qv_singleq}, the quadrupolar order with $\bm{q}=\QW$ leaves one of every two Pr layers disordered, resulting in a residual entropy of half of $\ln 2$ per Pr atom.
The residual entropy may eventually be released at lower temperatures through either a successive phase transition into another ordered state or screening by the Kondo effect.

Another possibility is a multiple-$\bm{q}$ order. Previous theoretical studies of the effective model with NN quadrupolar interactions on an fcc lattice have suggested a triple-$\bm{q}$ order at the X point~\cite{Tsunetsugu2021,Hattori2023}, which is represented by a linear combination of three configurations with different wave vectors $\QXa = (2\pi,~0,~0)$, $\QXb=(0,~2\pi,~0)$, and $\QXc=(0,~0,~2\pi)$.
Similarly, the W-point order involves three independent configurations with vectors $\QWa = (2\pi,~\pi,~0)$, $\QWb=(0,~2\pi,~\pi)$, and $\QWc=(\pi,~0,~2\pi)$, and hence multiple-$\bm{q}$ order may be possible.
The triple-$\bm{q}$ state at the X point is stabilized by the third-order coupling in the free energy of the form
\begin{align}
      \label{eq:third-free}
       Q_1(\QXa) Q_2(\QXb) Q_3(\QXc),
\end{align}
where $Q_1$, $Q_2$, and $Q_3$ denote arbitrary quadrupolar components.
This coupling is allowed because $\QXa+\QXb+\QXc=\bm{G}$, where $\bm{G}$ is a reciprocal lattice vector. 
In contrast, the W-point wave vectors do not satisfy this condition, i.e., $\QWa+\QWb+\QWc \neq \bm{G}$.
Therefore, the mechanism leading to a multiple-$\bm{q}$ state at the W point is nontrivial and is left for future study.

\section{Summary}
\label{sec:summary}

We extended the DFT+DMFT-based analysis of multipolar fluctuations, so far applied to $4f^1$-electron materials, to $4f^2$-electron systems.
The central challenge lies in the difference between the bases: The Hund's-rule ground state consists of many-particle states, whereas the Green's-function approach relies on a single-particle basis.
We constructed the $4f^2$ CEF levels of the Hund's multiplet based on the $j$--$j$ coupling scheme within the $j=5/2$ subspace.
We then established a method for calculating the susceptibility in the many-particle basis, $\mathcal{X}_{M_1M_2,M_3M_4}$. The susceptibility in the single-particle basis, $\chi_{m_1m_2,m_3m_4}$, is computed in the standard manner using the Bethe-Salpeter equation and then converted into $\mathcal{X}_{M_1M_2,M_3M_4}$ by an $f^1 \to f^2$ mapping.
The mapping relies on a reduction of the many-body Hilbert space by neglecting high-energy multiplets and on solving the inverse problem of the $f^2 \to f^1$ transformation.

We applied this framework to \prcdni, which exhibits a phase transition at $T_{\mathrm{O}}=1.0~\mathrm{K}$ within the $\Gamma_3$ CEF doublet.
From the analysis of the quadrupolar fluctuations and the intersite interactions, we proposed a $Q_u$-type quadrupolar order at the W point with the translation vector $\QW=(2\pi,\pi,0)$.
The subleading fourth-NN interaction is essential for stabilizing the W-point order over the X-point order favored by the NN interaction.

The present method enables the DFT+DMFT analysis of multipolar fluctuations for general $4f^n$ configurations.
Applications to other $4f^n$-electron systems, such as Nd and Sm compounds, are thereby made possible.

\begin{acknowledgments}
We thank T. Onimaru and H. Kusunose for valuable discussions.
This work was supported by JSPS KAKENHI Grants No.~23H04869, No.~25K22013, and No.~26K00659.
Part of the computations in this work were performed using the facilities of the Supercomputer Center, the Institute for Solid State Physics, the University of Tokyo (No.~2025-Ca-0086, No.~2025-Cb-0049, and 2026-Ca-0142).
\end{acknowledgments}

\appendix

\section{Calculation of matrix elements $A_{mm',MM'}$}
\label{app:A-matrix}

We present details of the calculation of the matrix elements $A_{mm',MM'}$ defined in Eq.~\eqref{eq:A-matrix}.
We first quote Eq.~\eqref{eq:A-matrix} with the $4f^n$ state $\ket{M}$ expressed explicitly as
\begin{align}
      \label{eq:A-matrix-2}
      A_{mm', MM'} = \bra{f^n\alpha(LS)JM} f^{\dag}_{jm} f_{j'm'} \ket{f^n\alpha(LS)JM'},
\end{align}
where $J$ is the total angular momentum resulting from the coupling of $L$ and $S$.
The label $\alpha$ distinguishes states with the same angular momenta.
In the less-than-half case ($n \le 7$), we insert a complete set of $4f^{n-1}$ states and obtain
\begin{align}
      \label{eq:A-intermediate-LS}
      &A_{mm',MM'}= \notag \\
      &\sum_{\alpha'L'M_L'S'M_S'}  \bra{f^n\alpha(LS)JM}f^{\dag}_{jm}\ket{f^{n-1}\alpha'L'M_L'S'M_S'} \notag \\
    & \times\bra{f^{n-1}\alpha'L'M_L'S'M_S'}f_{j'm'}\ket{f^n\alpha(LS)JM'},
\end{align}
where $M_L'$ and $M_S'$ denote the $z$ components of the orbital and spin angular momenta $L'$ and $S'$, respectively, of the intermediate states.
To evaluate the matrix element of the annihilation operator $f_{jm}$ in Eq.~\eqref{eq:A-intermediate-LS}, we first decouple the total angular momenta $j$ and $J$ into their orbital and spin components as
\begin{align}
      \label{eq:A-annihilation-1}
       & \bra{f^{n-1}\alpha'L'M_L'S'M_S'}f_{jm}\ket{f^n\alpha(LS)JM} \notag              \\
       & =\sum_{m_l\sigma M_LM_S}\bra{jm}\ket{lm_ls\sigma}\bra{LM_LSM_S}\ket{JM} \notag \\
       & \times\bra{f^{n-1}\alpha'L'M_L'S'M_S'}f_{lm_ls\sigma}\ket{f^nLM_LSM_S}.
\end{align}
The matrix elements of $f_{lm_ls\sigma}$ in the $LS$ basis can be evaluated using the coefficient of fractional parentage (cfp)~\cite{Judd1967,Otsuki2005-cfp}, given by
\begin{align}
      \label{eq:A-annihilation-2}
      \bra{f^{n-1}\alpha'L'M_L'S'M_S'}f_{lm_ls\sigma}\ket{f^nLM_LSM_S} \notag \\
      =(-1)^{n-1}\sqrt{n}
      (f^n\alpha LS\left\{|f^{n-1}\alpha'L'S',f\right.) \notag               \\
      \times\bra{L'M_L'lm_l}\ket{LM_L}\bra{S'M_S's\sigma}\ket{SM_S}.
\end{align}
Here, $(f^n\alpha LS\left\{|f^{n-1}\alpha'L'S',f\right.)$ denotes the cfp.

In the more-than-half case ($n \ge 8$), we use a hole picture.
Under the particle-hole transformation $h_m=f^{\dag}_m$ and $h^{\dag}_m=f_m$, $A_{mm',MM'}$ in Eq.~\eqref{eq:A-matrix} is rewritten as
\begin{align}
    \label{eq:A-matele-hole}
    A_{mm',MM'}=\delta_{mm'}\delta_{MM'}-\bra{M}h^{\dag}_{m'}h_{m}\ket{M'}
\end{align}
where $h^{\dag}\;(h)$ is the creation (annihilation) operator for a hole of $f$ orbitals and satisfies the anticommutation relation $\{h^{\dag}_m,h_{m'}\}=\delta_{mm'}$.
The $4f^n$ electron state $\ket{M}$ is now regarded as a $4f^{n_h}$ hole state with $n_h=14-n$ being the number of holes.
Therefore, Eq.~\eqref{eq:A-matele-hole} can be evaluated following the same procedure as in the less-than-half case [Eqs.~\eqref{eq:A-matrix-2}--\eqref{eq:A-annihilation-2}].
We note that the intermediate $4f^{n_h-1}$ hole states correspond to $4f^{n+1}$ electron states.

A complete list of cfp values is presented in Ref.~\cite{Nielson-cfp}. Table~\ref{tab:cfp-value} lists the values relevant to our calculations for the Hund’s-rule ground states. They satisfy the sum rule
\begin{align}
      \label{eq:cfp-sum}
    \sum_{\alpha'L'S'}(f^n\alpha LS\left\{|f^{n-1}\alpha'L'S',f\right.)^2=1.
\end{align}

As an example, we consider the $4f^2$ configuration. The many-particle states correspond to the Hund's-rule ground-state multiplet with $J=4$. Since the intermediate $4f^1$ state is uniquely determined by $l=3$ and $s=1/2$, the cfp simplifies to $(f^2\alpha LS\left\{|f^1ls,f\right.)=1$. The matrix elements $A_{mm',MM'}$ then take the form
\begin{align}
      \label{eq:Af2-intermediate}
      A_{mm',MM'}=
       & \sum_{m_l'\sigma'}\bra{f^2\alpha(LS)JM}f^{\dag}_{jm}\ket{f^1lm_l's\sigma'} \notag \\
       & \times\bra{f^1lm_l's\sigma'}f_{j'm'}\ket{f^2\alpha(LS)JM'}.
\end{align}
The matrix element of the annihilation operator $f_{jm}$ is given by
\begin{align}
      \label{eq:Af2-annihilation}
       & \bra{f^1lm_l's\sigma'}f_{jm}\ket{f^2\alpha(LS)JM} \notag    \\
       & = -\sqrt{2}\sum_{m_l\sigma M_LM_S}
      \bra{jm}\ket{lm_ls\sigma}\bra{lm_l'lm_l}\ket{LM_L} \notag      \\
       & \times\bra{s\sigma's\sigma}\ket{SM_S}\bra{LM_LSM_S}\ket{JM}.
\end{align}

\begin{table*}[!t]
\caption{Values of coefficients of fractional parentage (cfp) for general $4f^n$ electron configurations. The first, second, and third columns show the Hund's-rule configuration and the values of the total orbital and spin angular momenta $L$ and $S$ for the initial state $4f^n$. The fourth, fifth, and sixth columns specify the intermediate state, its possible electron configuration, and the corresponding $L'$ and $S'$ values, and the seventh column gives the corresponding cfp values, $(f^n\alpha LS\left\{|f^{n-1}\alpha'L'S',f\right.)$, taken from Ref.~\cite{Nielson-cfp}. Note that the intermediate states are $4f^{n-1}$ for the less-than-half cases ($n\le7$) and $4f^{n+1}$ for the more-than-half cases ($n\ge8$).}
\label{tab:cfp-value}
\centering
    \begin{tabular}{cccc|cccc|c}\hline
    $4f^n$ & multiplet & $L$ & $S$ & intermediate & multiplet & $L'$ & $S'$ & cfp \\\hline\hline
    $4f^2$(Pr$^{3+}$) & $^3H_{4}$ & $5$ & $1$ & $4f^1$ & $^2F_{5/2}$ & $3$ & $1/2$ & $1$ \\\hline
    $4f^3$(Nd$^{3+}$) & $^4I_{9/2}$ & $6$ & $3/2$ & $4f^2$ & $^3F_{2}$ & $3$ & $1$ & $-\sqrt{\frac{2}{9}}$ \\
        & & & & & $^3H_{4}$ & $5$ & $1$ & $\sqrt{\frac{7}{9}}$ \\\hline
    $4f^4$(Pm$^{3+}$) & $^5I_{4}$ & $6$ & $2$ & $4f^3$ & $^4F_{3/2}$ & $3$ & $3/2$ & $-\sqrt{\frac{1}{8}}$ \\
        & & & & & $^4G_{5/2}$ & $4$ & $3/2$ & $\sqrt{\frac{21}{88}}$ \\
        & & & & & $^4I_{9/2}$ & $6$ & $3/2$ & $\sqrt{\frac{7}{11}}$ \\\hline
    $4f^5$(Sm$^{3+}$) & $^6H_{5/2}$ & $5$ & $5/2$ & $4f^4$ & $^5D_{0}$ & $2$ & $2$ & $-\sqrt{\frac{2}{21}}$ \\
        & & & & & $^5F_{1}$ & $3$ & $2$ & $-\sqrt{\frac{1}{10}}$ \\
        & & & & & $^5G_{2}$ & $4$ & $2$ & $-\sqrt{\frac{39}{154}}$ \\
        & & & & & $^5I_{4}$ & $6$ & $2$ & $\sqrt{\frac{91}{165}}$ \\\hline
    $4f^6$(Eu$^{3+}$) & $^7F_{0}$ & $3$ & $3$ & $4f^5$ & $^6P_{1/2}$ & $2$ & $5/2$ & $\sqrt{\frac{1}{7}}$ \\
        & & & & & $^6F_{1/2}$ & $3$ & $5/2$ & $\sqrt{\frac{1}{3}}$ \\
        & & & & & $^6H_{5/2}$ & $5$ & $5/2$ & $\sqrt{\frac{11}{21}}$ \\\hline
    $4f^7$(Gd$^{3+}$) & $^8S_{7/2}$ & $0$ & $7/2$ & $4f^6$ & $^7F_{0}$ & $3$ & $3$ & $1$ \\\hline
    $4f^8$(Tb$^{3+}$) & $^7F_{6}$  & $3$ & $3$ & $4f^9$ & $^6P_{9/2}$ & $2$ & $5/2$ & $\sqrt{\frac{1}{7}}$ \\
        & & & & & $^6F_{11/2}$ & $3$ & $5/2$ & $\sqrt{\frac{1}{3}}$ \\
        & & & & & $^6H_{15/2}$ & $5$ & $5/2$ & $\sqrt{\frac{11}{21}}$ \\\hline
    $4f^9$(Dy$^{3+}$) & $^6H_{15/2}$ & $5$ & $5/2$ & $4f^{10}$ & $^5D_{4}$ & $2$ & $2$ & $-\sqrt{\frac{2}{21}}$ \\
        & & & & & $^5F_{7}$ & $3$ & $2$ & $-\sqrt{\frac{1}{10}}$ \\
        & & & & & $^5G_{6}$ & $4$ & $2$ & $-\sqrt{\frac{39}{154}}$ \\
        & & & & & $^5I_{8}$ & $6$ & $2$ & $\sqrt{\frac{91}{165}}$ \\\hline
    $4f^{10}$(Ho$^{3+}$) & $^5I_{8}$ & $6$ & $2$ & $4f^{11}$ & $^4F_{9/2}$ & $3$ & $3/2$ & $-\sqrt{\frac{1}{8}}$ \\
        & & & & & $^4G_{11/2}$ & $4$ & $3/2$ & $\sqrt{\frac{21}{88}}$ \\
        & & & & & $^4I_{15/2}$ & $6$ & $3/2$ & $\sqrt{\frac{7}{11}}$ \\\hline
    $4f^{11}$(Er$^{3+}$) & $^4I_{15/2}$ & $6$ & $3/2$ & $4f^{12}$ & $^3F_{4}$ & $3$ & $1$ & $-\sqrt{\frac{2}{9}}$ \\
        & & & & & $^3H_{6}$ & $5$ & $1$ & $\sqrt{\frac{7}{9}}$ \\\hline
    $4f^{12}$(Tm$^{3+}$) & $^3H_{6}$ & $5$ & $1$ & $4f^{13}$ & $^2F_{7/2}$ & $3$ & $1/2$ & $1$ \\\hline
    \end{tabular}
\end{table*}

\section{Overdetermined and underdetermined problems in general $4f^n$ configurations}
\label{app:over-under}

In Sec.~\ref{sec:fn-fluctuation}, we discussed the mapping of the susceptibility from the single-particle basis of $4f^1$ to the many-body basis of $4f^n$. This mapping is either an overdetermined or underdetermined problem, depending on the rare-earth ion. The $4f^2$ (Pr$^{3+}$) case was discussed in detail in Sec.~\ref{sec:fn-fluctuation}. Here, we consider general $4f^n$ cases.

Table~\ref{tab:Amatrix-ou} summarizes the conditions for rare-earth ions with configurations from $4f^2$ to $4f^{12}$.
We find that the overdetermined condition is realized for $4f^5$ (Sm$^{3+}$) and $4f^6$ (Eu$^{3+}$). Since Eu$^{3+}$ with $J=0$ is trivial, Sm$^{3+}$ offers an interesting test case for the overdetermined condition.
The other ions correspond to the underdetermined condition. In these cases, it is helpful to consider a reduction of the many-body basis based on the CEF splitting. In particular, ions with $n \geq 8$ have large $J$, and restricting the many-body basis to the low-energy states is especially useful.

We remark that the inclusion of the high-energy single-particle states of $4f^1$ ($j=7/2$ for $n\leq 7$ and $j=5/2$ for $n\geq 8$) does not change an underdetermined problem into an overdetermined one. The rank of the matrix $A$ is upper-bounded by 64, which corresponds to the number of operators within the $j=7/2$ subspace. This indicates that the $j=5/2$ and $j=7/2$ subspaces do not provide independent equations in the linear system.

\begin{table}[!t]
\caption{Characteristics of $A_{mm',MM'}$ for various $4f^n$ electron configurations. The first and second columns are the electron configuration of trivalent lanthanide ions, $4f^n$, and the total angular momentum $J$ of the corresponding Hund's-rule ground state, respectively. The third column ``$j(4f^1)$'' shows the $4f^1$ subspace considered. Here, $5/2$ and $7/2$ indicate the single-electron total angular momentum, and ``both" means that both $j=5/2$ and $7/2$ states are included. The fourth and fifth columns show the size of the matrix $A_{mm',MM'}$ and its rank, respectively. The column ``o/u" indicates whether the $4f^1 \to 4f^n$ problem corresponds to an overdetermined (over) or underdetermined (under) problem.}
\label{tab:Amatrix-ou}
    \begin{tabular}{cccccc}\hline
    $4f^n$ & $J(4f^n)$ & $j(4f^1)$ & Size of $A$ & rank($A$) & o/u \\\hline\hline
    $4f^2$(Pr$^{3+}$) & $4$ & $5/2$ & $36\times 81$ & $36$ & under \\
        & & both & $196\times 81$ & $64$ & under \\\hline
    $4f^3$(Nd$^{3+}$)& $9/2$ & $5/2$ & $36\times 100$ & $36$ & under \\
        & & both & $196\times 100$ & $64$ & under \\\hline
    $4f^4$(Pm$^{3+}$) & $4$  & $5/2$ & $36\times 81$ & $36$ & under \\
        & & both & $196\times 81$ & $64$ & under \\\hline
    $4f^5$(Sm$^{3+}$) & $5/2$ & $5/2$ & $36\times 36$ & $36$ & over \\
        & & both & $196\times 36$ & $36$ & over \\\hline
    $4f^6$(Eu$^{3+}$) & $0$ & $5/2$ & $36\times 1$ & $1$ & over \\
        & & both & $196\times 1$ & $1$ & over \\\hline
    $4f^7$(Gd$^{3+}$) & $7/2$ & $7/2$ & $64\times 64$ & $4$ & under \\
        & & both & $196\times 64$ & $4$ & under \\\hline
    $4f^8$(Tb$^{3+}$) & $6$ & $7/2$ & $64\times 169$ & $49$ & under \\
        & & both & $196\times 169$ & $49$ & under \\\hline
    $4f^9$(Dy$^{3+}$) & $15/2$ & $7/2$ & $64\times 256$ & $64$ & under \\
        & & both & $196\times 256$ & $64$ & under \\\hline
    $4f^{10}$(Ho$^{3+}$) & $8$ & $7/2$ & $64\times 289$ & $64$ & under \\
        & & both & $196\times 289$ & $64$ & under \\\hline
    $4f^{11}$(Er$^{3+}$) & $15/2$ & $7/2$ & $64\times 256$ & $64$ & under \\
        & & both  & $196\times 256$ & $64$ & under \\\hline
    $4f^{12}$(Tm$^{3+}$) & $6$ & $7/2$ & $64\times 169$ & $64$  & under \\
        & & both & $196\times 169$ & $64$ & under \\\hline
    \end{tabular}
\end{table}

\section{Adjusting CEF parameters $w$ and $x$}
\label{app:adjusting-wx}

We present a procedure for adjusting the CEF parameters $w$ and $x$ to reproduce the experimentally estimated $4f^2$ ($J=4$) CEF levels of \prcdni.
In Sec.~\ref{sec:CEF}, the CEF splitting within the $J=4$ multiplet is reproduced from the eigenvalues of a many-particle Hamiltonian that incorporates Coulomb interactions into the local $f$-electron Hamiltonian.
Since the CEF parameters $w$ and $x$ in Eq.~\eqref{eq:f1-CEF} are not equivalent to $W$ and $X$ in Eq.~\eqref{eq:f2-CEF}, the experimental values $W=-3\;\mathrm{K}$ and $X=0.8$ do not yield a $\Gamma_3$ doublet ground state when employed in the $j$--$j$ coupling scheme.

We first determine the scale parameter $w$. To this end, we introduce a measure of the CEF splitting as
\begin{align}
\label{eq:scale-CEF}
\dEjj(w,x)\equiv \sqrt{\overline{E_i(w,x)^2}-\left(\overline{E_i(w,x)}\right)^2},
\end{align}
where $E_i(w,x)$ ($i=1,\ldots,9$) denote the CEF eigenvalues within the $J=4$ multiplet calculated using the $j$--$j$ coupling scheme, and the overline denotes an average over these levels.
Similarly, we define the corresponding measure of the CEF splitting evaluated within the $LS$ coupling scheme as $\dELS(W,X)$.
We evaluate $\dEjj(w,x)$ using a trial value $w=w^{(0)}$ and $x=X$. The ratio between $\dELS$ and $\dEjj$ then yields the appropriate scale $w$ as
\begin{align}
\label{eq:ratio}
w=w^{(0)}\frac{\dELS(W,X)}{\dEjj(w^{(0)},X)}.
\end{align}
With this value of $w$, the CEF levels are evaluated by scanning $x$ and plotted as shown in Fig.~\ref{fig:LLW}(d).
We then choose $x$ so that the level scheme and the energy splitting between the ground state and the first-excited state agree with the experiment.
For \prcdni, we obtain $w=4.23\;\mathrm{K}$ and $x=0.62$, which reproduce the experimental value $\Delta_{\Gamma_3-\Gamma_4}=13.2\;\mathrm{K}$.

\section{Temperature dependence of the local susceptibility in the large CEF case}
\label{app:t-chiloc-10w}

\begin{figure}[!t]
      \includegraphics[width=1.0\linewidth]{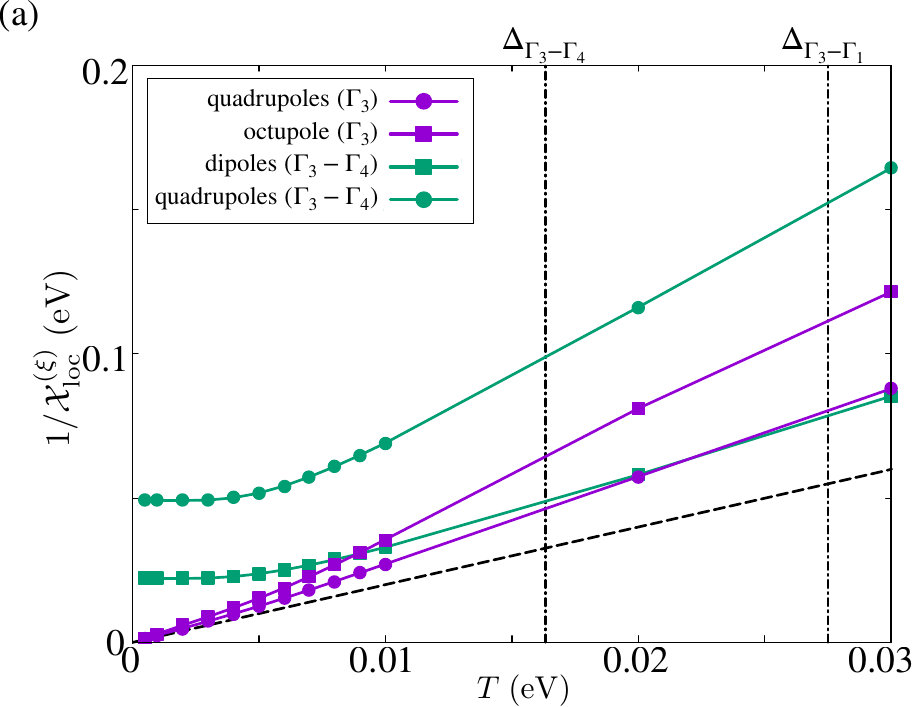}
      \includegraphics[width=1.0\linewidth]{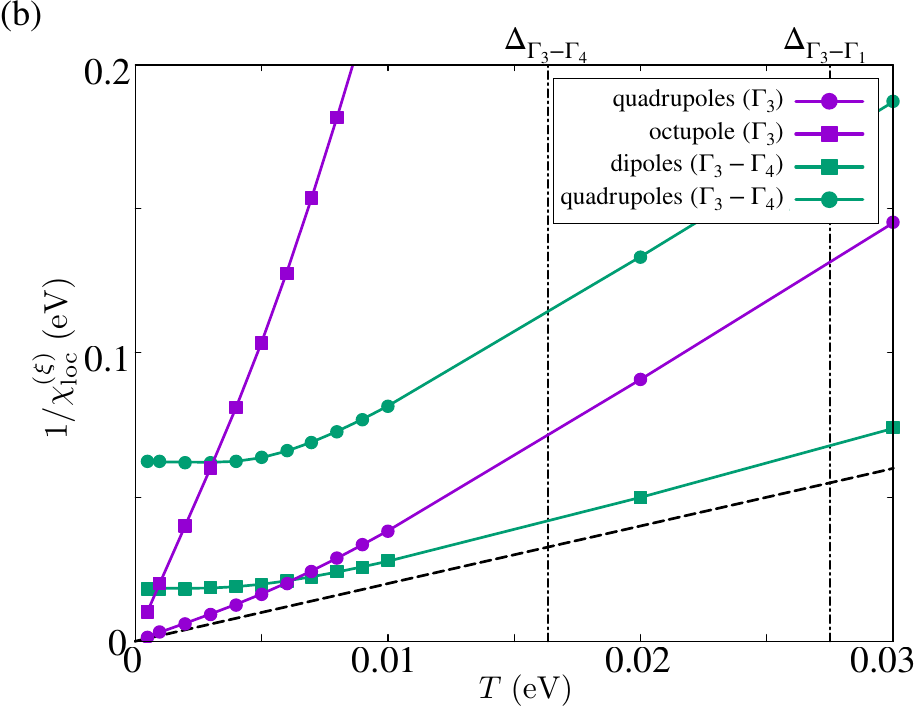}
      \caption{Temperature dependence of the inverse of the local susceptibilities in the large CEF case: (a) $\mathcal{X}_{\mathrm{loc}}^{(\xi)}$ for the many-particle basis and (b) $\chi_{\mathrm{loc}}^{(\xi)}$ for the single-particle basis. The dashed line shows the Curie law at low $T$. The vertical dash-dotted lines indicate the energies of the CEF splitting between the ground state $\Gamma_3$ and the excited states $\Gamma_4$ and $\Gamma_1$.}
      \label{fig:t-chiinv_10w}
\end{figure}

The results in the main text are calculated using the realistic CEF level scheme.
The low-lying CEF states are separated by $\Delta_{\Gamma_3-\Gamma_4} \approx 14\,\mathrm{K}$, for which the lowest temperature computed, $T=0.5\,\mathrm{meV}$, is not sufficiently low.
In this Appendix, we present results for a larger CEF scale, which allows us to access a sufficiently low-temperature regime.
By enlarging the CEF parameter as $w\to10w$, we obtain $\Delta_{\Gamma_3-\Gamma_4}\approx 189\;\mathrm{K}$, $\Delta_{\Gamma_3-\Gamma_1}\approx319\;\mathrm{K}$, and $\Delta_{\Gamma_3-\Gamma_5}\approx1199\;\mathrm{K}$.

Figure~\ref{fig:t-chiinv_10w}(a) shows the temperature dependence of the eigenvalues $\mathcal{X}_{\mathrm{loc}}^{(\xi)}$ of the local susceptibility corresponding to Fig.~\ref{fig:t-chiinv}, with the CEF parameter scaled as $w\to10w$.
It is now clear that the fluctuations within the $\Gamma_3$ doublet (the electric quadrupoles and the magnetic octupole) approach the Curie law [Eq.~\eqref{eq:curie}] with the Curie constant $C=1/2$, and that the fluctuations involving $\Gamma_3$ and the first-excited $\Gamma_4$ states (the magnetic dipoles and the electric quadrupoles) are suppressed at low temperatures.

$\mathcal{X}_{\mathrm{loc}}^{(\xi)}$ in Fig.~\ref{fig:t-chiinv_10w}(a) is obtained through the mapping from $4f^1$ to $4f^2$.
Figure~\ref{fig:t-chiinv_10w}(b) shows the susceptibility before the mapping, namely, $\chi_{\mathrm{loc}}^{(\xi)}$
in the $4f^1$ basis.
The main difference compared with the $4f^2$ counterpart [Fig.~\ref{fig:t-chiinv_10w}(a)] is found in the octupole component.
It follows the Curie law as expected, but with a Curie constant $C \approx 0.05$, which is ten times smaller than the expected value $C=1/2$.
This deviation reflects the weight of the $4f^1$ octupole moment in the CEF ground state of the $4f^2$ configuration.
This demonstrates the utility of the mapping from $4f^1$ to $4f^2$ in recovering the physical value of the Curie constant.

\bibliography{refs}

\end{document}